\documentclass[aps,prc,showpacs,twocolumn,superscriptaddress,10pt]{revtex4}
\usepackage{amsmath} % Include Math and Symbols
\usepackage{amssymb}
\usepackage{graphicx}% Include figure files
\usepackage{graphics}
\usepackage{dcolumn}% Align table columns on decimal point
\usepackage{bm}% bold math
\usepackage{epsfig}
\usepackage{epstopdf}
\usepackage{multirow}
\usepackage{color}
\begin{document}
\title{Strange and Multi-strange Particle Production in Au+Au Collisions at $\sqrt{s_{NN}}=62.4$~GeV}
%--------------------------------------------------

%with Forcing affiliation order
 \affiliation{Argonne National Laboratory, Argonne, Illinois 60439, USA}
\affiliation{Brookhaven National Laboratory, Upton, New York 11973, USA}
\affiliation{University of California, Berkeley, California 94720, USA}
\affiliation{University of California, Davis, California 95616, USA}
\affiliation{University of California, Los Angeles, California 90095, USA}
\affiliation{Universidade Estadual de Campinas, Sao Paulo, Brazil}
\affiliation{University of Illinois at Chicago, Chicago, Illinois 60607, USA}
\affiliation{Creighton University, Omaha, Nebraska 68178, USA}
\affiliation{Czech Technical University in Prague, FNSPE, Prague, 115 19, Czech Republic}
\affiliation{Nuclear Physics Institute AS CR, 250 68 \v{R}e\v{z}/Prague, Czech Republic}
\affiliation{University of Frankfurt, Frankfurt, Germany}
\affiliation{Institute of Physics, Bhubaneswar 751005, India}
\affiliation{Indian Institute of Technology, Mumbai, India}
\affiliation{Indiana University, Bloomington, Indiana 47408, USA}
\affiliation{Alikhanov Institute for Theoretical and Experimental Physics, Moscow, Russia}
\affiliation{University of Jammu, Jammu 180001, India}
\affiliation{Joint Institute for Nuclear Research, Dubna, 141 980, Russia}
\affiliation{Kent State University, Kent, Ohio 44242, USA}
\affiliation{University of Kentucky, Lexington, Kentucky, 40506-0055, USA}
\affiliation{Institute of Modern Physics, Lanzhou, China}
\affiliation{Lawrence Berkeley National Laboratory, Berkeley, California 94720, USA}
\affiliation{Massachusetts Institute of Technology, Cambridge, MA 02139-4307, USA}
\affiliation{Max-Planck-Institut f\"ur Physik, Munich, Germany}
\affiliation{Michigan State University, East Lansing, Michigan 48824, USA}
\affiliation{Moscow Engineering Physics Institute, Moscow Russia}
\affiliation{NIKHEF and Utrecht University, Amsterdam, The Netherlands}
\affiliation{Ohio State University, Columbus, Ohio 43210, USA}
\affiliation{Old Dominion University, Norfolk, VA, 23529, USA}
\affiliation{Panjab University, Chandigarh 160014, India}
\affiliation{Pennsylvania State University, University Park, Pennsylvania 16802, USA}
\affiliation{Institute of High Energy Physics, Protvino, Russia}
\affiliation{Purdue University, West Lafayette, Indiana 47907, USA}
\affiliation{Pusan National University, Pusan, Republic of Korea}
\affiliation{University of Rajasthan, Jaipur 302004, India}
\affiliation{Rice University, Houston, Texas 77251, USA}
\affiliation{Universidade de Sao Paulo, Sao Paulo, Brazil}
\affiliation{University of Science \& Technology of China, Hefei 230026, China}
\affiliation{Shandong University, Jinan, Shandong 250100, China}
\affiliation{Shanghai Institute of Applied Physics, Shanghai 201800, China}
\affiliation{SUBATECH, Nantes, France}
\affiliation{Texas A\&M University, College Station, Texas 77843, USA}
\affiliation{University of Texas, Austin, Texas 78712, USA}
\affiliation{Tsinghua University, Beijing 100084, China}
\affiliation{United States Naval Academy, Annapolis, MD 21402, USA}
\affiliation{Valparaiso University, Valparaiso, Indiana 46383, USA}
\affiliation{Variable Energy Cyclotron Centre, Kolkata 700064, India}
\affiliation{Warsaw University of Technology, Warsaw, Poland}
\affiliation{University of Washington, Seattle, Washington 98195, USA}
\affiliation{Wayne State University, Detroit, Michigan 48201, USA}
\affiliation{Institute of Particle Physics, CCNU (HZNU), Wuhan 430079, China}
\affiliation{Yale University, New Haven, Connecticut 06520, USA}
\affiliation{University of Zagreb, Zagreb, HR-10002, Croatia}

\author{M.~M.~Aggarwal}\affiliation{Panjab University, Chandigarh 160014, India}
\author{Z.~Ahammed}\affiliation{Lawrence Berkeley National Laboratory, Berkeley, California 94720, USA}
\author{A.~V.~Alakhverdyants}\affiliation{Joint Institute for Nuclear Research, Dubna, 141 980, Russia}
\author{I.~Alekseev~~}\affiliation{Alikhanov Institute for Theoretical and Experimental Physics, Moscow, Russia}
\author{J.~Alford}\affiliation{Kent State University, Kent, Ohio 44242, USA}
\author{B.~D.~Anderson}\affiliation{Kent State University, Kent, Ohio 44242, USA}
\author{C.~D.~Anson}\affiliation{Ohio State University, Columbus, Ohio 43210, USA}
\author{D.~Arkhipkin}\affiliation{Brookhaven National Laboratory, Upton, New York 11973, USA}
\author{G.~S.~Averichev}\affiliation{Joint Institute for Nuclear Research, Dubna, 141 980, Russia}
\author{J.~Balewski}\affiliation{Massachusetts Institute of Technology, Cambridge, MA 02139-4307, USA}
% ...

\collaboration{STAR Collaboration}\noaffiliation

%==================================================
\begin{abstract}
We present results on strange and multi-strange particle production in Au+Au collisions at $\sqrt{s_{NN}}=62.4$~GeV as measured with the STAR detector at RHIC. 
Mid-rapidity transverse momentum spectra and integrated yields of $K^{0}_{S}$, $\Lambda$, $\Xi$, $\Omega$ and their anti-particles are presented for different centrality classes. 
The particle yields and ratios follow a smooth energy dependence. 
Chemical freeze-out parameters, temperature, baryon chemical potential and strangeness saturation factor obtained from the particle yields are presented.
Intermediate transverse momentum ($p_T$) phenomena are discussed based on the ratio of the measured baryon-to-meson spectra and nuclear modification factor. 
The centrality dependence of various measurements presented show a similar behavior as seen in Au+Au collisions at $\sqrt{s_{NN}}=200$~GeV.
\end{abstract}

%\keywords{}
\maketitle

\section{Introduction}
\label{lab:Introduction}
%--------------------------------------------------
%==================================================
Strange hadrons in the final products of high-energy nuclear collisions provide valuable insight into the properties of the created system, since they are not present inside the nuclei of the incoming beams. 
The production rates and phase space distributions of the strange particles may reveal different characteristics of the created fireball and thus they have been studied at different accelerators and experiments~\cite{ref:E896_Caines_JPG27_2001_311,ref:NA57_PLB595_2004_68,ref:NA49_PRL,ref:STAR_PRL89_2002_092301}.

The enhancement of strangeness production in relativistic heavy ion collisions was first proposed as a signature of the Quark Gluon Plasma (QGP) in the early 1980s~\cite{ref:Rafelski_2}. 
The observation of an enhancement of strange baryon production relative to $p+p$ collisions in SPS data~\cite{ref:NA35,ref:NA57,ref:WA85}, confirmed later by data from the Relativistic Heavy Ion Collider (RHIC)~\cite{ref:HelenSTAR}, has brought a lot of excitement to these studies.

While the production yields are mainly determined by the yields in the low transverse momentum region ($p_T < 2$~GeV/c), the intermediate transverse momentum region ($p_{T}$ between 2 and 5 GeV/c) might give insight into the hadron production mechanism which in turn reveals important characteristics of the system. 
One of the most interesting results observed in this region is the apparent enhancement of baryons over mesons which was originally observed in the $p/\pi$ ratio~\cite{ref:STAR_PLB655_2007_104} but was also confirmed in the strangeness sector with the $\Lambda/K^{0}_{S}$ ratio~\cite{ref:ProccedingMatt}.
Different theoretical scenarios to explain such behavior are discussed showing that detailed experimental data are needed to discriminate between these proposals.

We present in this paper results on the production of $K^{0}_{S}$, $\Lambda$, $\Xi$ and $\Omega$ obtained by the STAR experiment at RHIC in Au+Au collisions at an energy of $\sqrt{s_{NN}}=62.4$~GeV. 
This energy lies in the range between SPS and the highest RHIC energy and provides intermediate measurement points in the energy excitation functions. 
The systematic study with detailed measurement of the excitation function of various observables is very important for understanding the QCD phase diagram of nuclear matter and for the search for a critical point of the phase transition that would correspond to the end point of the first order phase boundary as predicted by QCD~\cite{ref:CriticalPoint1,ref:CriticalPoint2}.

The paper is organized as follows.
Section \ref{sec:Experiment} is devoted to the STAR experimental apparatus used for the measurements and the event selection. 
The analysis techniques to reconstruct the decays of the strange particles are described in section \ref{sec:analysis} along with the corrections needed to extract the physical spectra, the functional forms used to fit and integrate the $p_{T}$ spectra, and the procedure used to calculate the systematic uncertainties. 
In section \ref{sec:results} the main results are presented and the physical implications of these measurements are discussed, including detailed comparisons to the corresponding data from Au+Au at $200$~GeV collisions \cite{ref:HelenSTAR, ref:Fuqiang, ref:STAR_PRL98_2007_062301}.
Chemical freeze-out properties are extracted from the particle ratios, strangeness enhancement is tested using the ratio of strange particles and pions, and baryon-to-meson differences are studied by comparing the $K^{0}_{S}$ spectra to the $\Lambda$ spectra.
Final remarks and an overview of the general behavior of strange particle production considering the collision energy dependence are presented in section \ref{sec:conclusion}.

%==================================================
\section{Experimental Apparatus}
\label{sec:Experiment}
%--------------------------------------------------
%==================================================
The data presented in this paper were measured during the 2004 run of RHIC situated at Brookhaven National Laboratory (BNL) using the STAR detector. 
The data correspond to Au+Au collisions at $\sqrt{s_{NN}}=62.4$~GeV. 
Descriptions and references concerning the STAR detector and its sub-detectors may be found in Ref.~\cite{ref:STARNIM}. 
The main subsystem of the STAR detector used for this analysis is a cylindrical time projection chamber (TPC) (length 4.2 m, inner radius 0.5 m, outer radius 2 m) covering a pseudo-rapidity range of $|\eta|<1.8$ with a 2$\pi$ coverage in azimuth. 
The TPC is immersed in a magnetic field of 0.5 T parallel to the beam direction that enables the measurement of trajectories and momenta of charged particles with a transverse momentum ($p_{T}$) above 0.15 GeV/c. The TPC  is filled with P10 gas (10\% methane, 90\% argon) regulated at 2 mbar above atmospheric pressure.

In addition to its track detection and momentum determination capabilities, the TPC provides particle-identification for charged particles by measuring their ionization energy-loss ($dE/dx$) in the TPC gas. 
Details and performance of the energy-loss particle-identification method are explained in Ref.~\cite{ref:NIMPID}. 

The collisions were triggered using a combination of several trigger detectors~\cite{ref:STARTrigger}. 
The minimum bias trigger requires coincidences between two zero degree calorimeters (ZDC) located $18$~m along the beam axis on each side of the interaction region. Beam-fragment (spectator) neutrons are not deflected by the beam guiding magnets and are thus detected in the ZDC detectors. In addition, a scintillation counter central trigger barrel (CTB) that surrounds the outer cylinder of the TPC was used to trigger on charged particle multiplicity in the mid-pseudo-rapidity region. Two beam-beam counters (BBC) were also used in the trigger system to measure the charged particle multiplicity in the forward region ($3.3<|\eta|<5.0$). 

The acquired data consist of a total of $6\times 10^{6}$ minimum bias triggered events. 
In order to study the centrality dependence of strangeness production, these events were divided into centrality classes according to fractions of the total inelastic collision cross-section. 
The centrality of an event was defined as in previous analyses~\cite{ref:Fuqiang}, by the measured (uncorrected) multiplicity in the  TPC of charged tracks with specific quality cuts. 

In this analysis, only events with a primary vertex position within $\pm$30 cm from the center of the TPC along the beam line were used to ensure a good rapidity coverage and relative uniformity of the detector efficiency. 
All the results quoted in this paper are mid-rapidity ($|y|<1.0$) results. 
For each event centrality class, the equivalent collision impact parameter is calculated based on a comparison between the measured uncorrected charged particle multiplicity $dN_{ch}/d\eta$ and the calculated cross section using a Monte Carlo Glauber model~\cite{ref:Glauber}. 
With this comparison, it is possible to determine for each centrality class the average number of participant nucleons $\langle N_{\rm part} \rangle$ and the average number of nucleon-nucleon binary collisions $\langle N_{\rm bin} \rangle$. 
These numbers together with the mid-rapidity charged particle yields $dN_{ch}/dy$ are shown for all centrality classes in Table~\ref{tab:Centrality}.

%Table I
\begin{table*}[hbt]
\begin{center}
\caption{Summary of the centrality classes used to separate the events in this analysis. For each centrality class, the equivalent mean number of participant nucleons $\langle N_{\rm part} \rangle $ and the mean number of binary collisions are determined using the charged particle multiplicity distributions and a Monte Carlo Glauber model. 
Also shown in this table is the corrected charged particle multiplicity density $\langle dN_{\rm ch} / dy \rangle $ that was obtained by adding the yields of protons, anti-protons, charged pions and charged kaons. Details of the analysis can be found in Ref.~\cite{ref:Fuqiang}. }
\begin{tabular}{c|ccccccccccccc}
\hline\hline
 & $0-5$\% & \hspace{0.25cm} & $5-10$\% & \hspace{0.25cm} & $10-20$\% & \hspace{0.25cm} & $20-30$\% & \hspace{0.25cm} & $30-40$\% & \hspace{0.25cm} & $40-60$\% & \hspace{0.25cm} & $60-80$\% \\
\hline
\hline
\quad $\langle N_{\rm part} \rangle $ & 346.5 $\pm$ 2.8 & & 293.9 $\pm$ 4.2 & & 229.8 $\pm$ 4.6 & & 164.1 $\pm$ 5.4 & & 114.3 $\pm$ 5.1 & & 62.0  $\pm$ 4.9 & & 21.4  $\pm$ 3.0  \\
\hline
\quad $\langle N_{\rm bin} \rangle $ & 891 $\pm$ 57 & & 710 $\pm$ 47 & & 511 $\pm$ 34 & & 325 $\pm$ 23 & & 199 $\pm$ 16 & & 88 $\pm$ 10 & & 22 $\pm$ 4 \\
\hline
\quad $\langle dN_{\rm ch} / dy \rangle $ & 582 $\pm$ 38 & & 476 $\pm$ 30 & & 359 $\pm$ 24 & & 249 $\pm$ 16 & & 166 $\pm$ 11 & & 86 $\pm$ 7 & & 27 $\pm$ 2 \\
\hline
\hline
\end{tabular}
\label{tab:Centrality}
\end{center}
\end{table*}

%==================================================
\section{Analysis}
\label{sec:analysis}
%==================================================

% Fig01
\begin{figure*}[hbt]
\includegraphics[width=13cm]{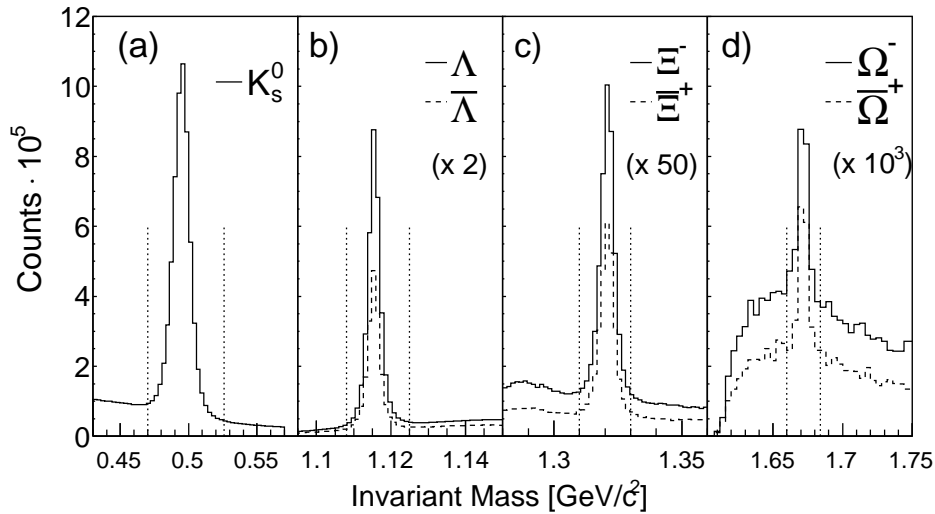}
\caption{Invariant mass distributions for selected candidates for (a) $K^{0}_{S}$, (b) $\Lambda$, $\overline{\Lambda}$, (c) $\Xi^{-}$, $\bar{\Xi}^{+}$ and (d) $\Omega^{-}$, $\bar{\Omega}^{+}$ (note the different scaling factors for $\Lambda$,  $\Xi$ and $\Omega$) for rapidity interval $|y|<1.0$ and $p_{T}>0.2$ ~GeV/c. The limits represented by the vertical dotted lines are the ones used for signal counting.}
\label{fig:invariantMasses}
\end{figure*}

The analysis methods for the different particle species are detailed in the following sub-sections. 
The $K^{0}_{S}$ and $\Lambda$ reconstruction method is described in \ref{subsec:V0_recon}. 
The reconstruction of multi-strange particles through the correlation of 3 particles is described in section~\ref{subsec:Xi_recon}.

%==================================================
\subsection{$K^{0}_{S}$ and $\Lambda$ reconstruction} 
\label{subsec:V0_recon}
%==================================================

$K^{0}_{S}$, $\Lambda$ and $\bar{\Lambda}$ were identified through the reconstruction of the weak decay topology: 
\vspace{0.15cm}
$$K^{0}_{S}  \rightarrow \pi^{+} + \pi^{-}\vspace{-0.25cm}$$
(branching ratio 69.20\%)
\vspace{0.15cm}
$$\Lambda  \rightarrow p + \pi^{-} \vspace{-0.5cm}$$
$$ \overline{\Lambda} \rightarrow \overline{p} + \pi^{+} \vspace{-0.25cm}$$
(branching ratio 63.9\%). 
\vspace{0.15cm}

In addition to the $dE/dx$ information of the daughter tracks, geometrical cuts such as the distance of closest approach (DCA) between the two daughter tracks and the pointing vector of the daughter tracks (distance of closest approach of the projected helix trajectory) away from the primary vertex position were used to improve the signal to noise ratio of the invariant mass peaks. 
A geometrical cut on the extrapolation of the reconstructed $\Lambda$ trajectory from the primary vertex position was also used to reduce the contribution of feed-down $\Lambda$ particles such as those  from the $\Xi$ sequential decay. 
Cut selection was optimized considering a compromise between background minimization and signal efficiency in the high-$p_T$ bins. Table \ref{tab:cuts_V0} shows the geometrical cuts used in this analysis.

%Table II
\begin{table*}[hbt]
\begin{center}
\caption{Geometrical cuts used in the analysis of $K_S^0$ and $\Lambda$($\bar{\Lambda}$) production.}
\begin{tabular}{c|c|c}
\hline
\hline
\begin{minipage}[c][0.5cm][c]{0.33\columnwidth} Cut \end{minipage} & \begin{minipage}[c][0.5cm][c]{0.33\columnwidth} $K_s^0$ \end{minipage} & \begin{minipage}[c][0.5cm][c]{0.33\columnwidth} $\Lambda$ and $\bar{\Lambda}$ \end{minipage}\\
\hline
\hline
DCA of V0 to primary vertex & $ < 0.5$ cm & $< 0.8$ cm \\
DCA of V0-daughters to primary vertex & $> 0.8$ cm & $> 1.0$ cm \\
DCA between V0-daughters & $< 0.8$ cm & $< 0.8$ cm \\
Number of hits in the daughters trajectory & $\ge 15$ & $\ge 15$ \\
Radial decay length & $> 4$ cm & $>5$ cm \\
\hline
\hline
\end{tabular}
\label{tab:cuts_V0}
\end{center}
\end{table*}

The invariant mass distributions of the reconstructed particles, obtained after the geometrical and particle identification cuts, are presented in Fig.~\ref{fig:invariantMasses} panel (a) for $K^{0}_{S}$ and panel (b) for $\Lambda$ and $\overline{\Lambda}$. 
The remaining background underneath the mass peak was subtracted by making an interpolation of the spectrum on either side of the mass peak. 
Various polynomial functions were fitted to the background to estimate a systematic uncertainty for the signal area.
The difference in the background estimated using the two different methods was used in the calculation of the systematic uncertainty for each point in the final $p_{T}$ spectra. 
In general, the shape of the background in the invariant mass spectra was smooth and relatively flat which resulted in uncertainties below 3\%. 
The signal itself was then determined by counting the entries in the bins contributing to the peak and subtracting the estimated background.

%==================================================
\subsection{$\Xi$ and $\Omega$ reconstruction}
\label{subsec:Xi_recon}
%==================================================

The multi-strange hyperons were reconstructed via the topology of their weak decays:
\vspace{-0.25cm}
$$\Xi^- \rightarrow \Lambda + \pi^- \vspace{-0.5cm}$$
$$\bar{\Xi}^{+} \rightarrow \overline{\Lambda}  + \pi^{+}\vspace{-0.25cm}$$ (branching ratio 99.887\%),
\vspace{-0.25cm}
 $$\Omega^-  \rightarrow \Lambda + \mbox{K}^- \vspace{-0.5cm}$$
$$\bar{\Omega}^{+} \rightarrow \overline{\Lambda}  + \mbox{K}^{+} \vspace{-0.25cm}$$ (branching ratio 67.8\%).

% Table III
\begin{table*}[hbt]
\begin{center}
\caption{Geometrical cuts used in the analysis of the production of $\Xi$, $\Omega$ and their anti-particles.}
\begin{tabular}{c|c|c}
\hline
\hline
 \begin{minipage}[c][0.5cm][c]{0.33\columnwidth} Cut \end{minipage} & \begin{minipage}[c][0.5cm][c]{0.33\columnwidth} $\Xi^{-}$ and $\bar{\Xi}^{+}$ \end{minipage} & \begin{minipage}[c][0.5cm][c]{0.33\columnwidth} $\Omega^-$ and $\bar{\Omega}^{+}$ \end{minipage}\\
\hline
\hline
DCA of parent to primary vertex & $ < 0.6$ cm & $< 0.6$ cm \\
DCA of bachelor to primary vertex & $ > 1.0$ cm & $>1.6\cdot \sqrt{\text{DCA of parent to primary vertex}}+0.1$ cm \\
DCA of V0 to primary vertex & $ > 0.5$ cm & $>1.4 \cdot \sqrt{\text{DCA of parent to primary vertex}}+0.1$ cm \\
DCA of positive V0-daughter to primary vertex & $> 0.7$ cm & - \\
DCA of negative V0-daughter to primary vertex & $> 3.0$ cm & - \\
DCA between V0 and bachelor & $< 0.8$ cm & $< 0.6$ cm \\
DCA between V0-daughters & $< 0.7$ cm & $< 0.6$ cm \\
Number of hits in the daughters trajectory & $\ge 15$ & $\ge 25$ \\
Parent decay length & $> 5$ cm & $>3$ and $<20$ cm \\
V0 decay length & - & $>2$ and $<30$ cm and $> -4*\sqrt{\text{$\Xi$ decay length}} + 23$ cm \\
Daughter V0 inv. mass & $1.1175 \pm 0.0125$ GeV$/c^2$ & $1.1150 \pm 0.0070$ GeV$/c^2$ \\
\hline
\hline
\end{tabular}
\label{tab:cuts_Xi}
\end{center}
\end{table*}

$\Lambda$ candidates reconstructed as described in the previous section are combined with single track assuming it to be the bachelor $\pi^{\pm}$ for the $\Xi$ decays and $K^{\pm}$ for the $\Omega$ decays.

Despite the identification of the daughters by their $dE/dx$, geometric selections were also used to improve the purity of the samples due to the large combinatorial background. 
These selections were done by cutting sequentially on the different variables aiming for a compromise between background minimization and signal efficiency in the high-$p_T$ bins. Table \ref{tab:cuts_Xi} shows the geometrical cuts used in this analysis. 
The invariant mass of the assumed parent particle obtained from the candidates passing the selection cuts was calculated and plotted in a histogram, as shown in panel (c) for the $\Xi$ and (d) for the $\Omega$ of Fig.~\ref{fig:invariantMasses}.

Even after the cut optimization, some background still remained in the invariant mass distributions. 
The background under the invariant mass peak was estimated using the same method as used for the $\Lambda$ analysis, where polynomial functions were used to parameterize the background outside the peak region.
The background estimated by a polynomial fit function was subtracted from the invariant mass spectra in order to calculate the raw yield. 
In the following step, the raw spectra were corrected for the detector and analysis inefficiencies.

%==================================================
\subsection{Correction Factors}
\label{subsec:CorrectionFactor}
%==================================================

\begin{figure}
 \includegraphics[width=8.5cm]{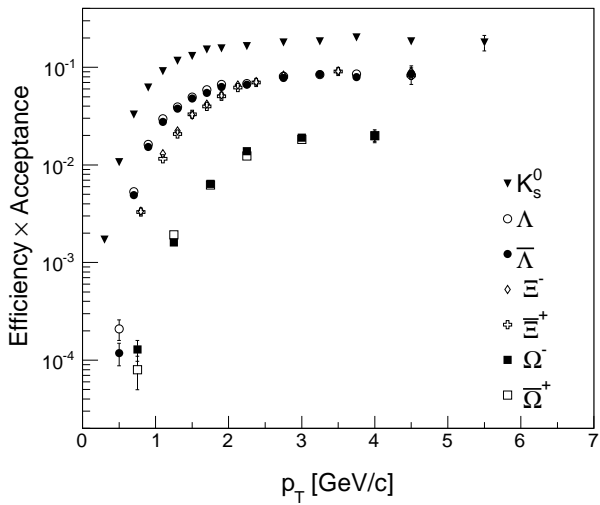}%
 \caption{Correction factors (acceptance $\times$ efficiency) for the most central events ( $0-5\%$ for $K^{0}_{S}$, $\Lambda$ and $\Xi$; $0-20\%$ for $\Omega$) at mid-rapidity ($|y|<1$) as a function of $p_T$ for the different particle species as obtained via embedding. The branching ratio of the measured decay channel is not factored into this plot.}
\label{fig:Efficiency}
\end{figure}

These corrections include tracking efficiency and detector acceptance as well as reconstruction inefficiencies. 
The correction factors were determined as a function of $p_T$ and the efficiencies for the different particles were calculated in different rapidity intervals between $-1$ and $+1$ to verify that these corrections were independent of rapidity.
The corrections were estimated by using Monte Carlo generated particles which were propagated through a TPC detector response simulator and embedded into real events on the cluster level. 
These enriched events were then reconstructed using the usual reconstruction chain. 
The correction factors for the different particles obtained by this method are shown as a function of $p_T$ in Fig.~\ref{fig:Efficiency}. 

% Fig02

The measured $\Lambda$ spectra were also corrected for feed-down of weak decays  by 
 subtracting the contributions from the charged and neutral $\Xi$ decays.
As already mentioned in section~\ref{subsec:V0_recon}, a geometrical cut on the distance of closest approach between the extrapolated reconstructed track of the $\Lambda$ and the primary vertex position was used to reduce this contribution. 
Nevertheless, some of the secondary $\Lambda$ still satisfied this cut, especially in the high-$p_{T}$ region. 
The relative number of $\Lambda$ particles from the $\Xi$ decays that were in the primary particle sample was estimated using 
embedded Monte Carlo $\Xi$ to calculate the reconstruction efficiency of these secondary $\Lambda$ particles which was then scaled by the corrected yield of the measured $\Xi$ particles. 
The relative contribution of the feed-down was calculated for each $p_{T}$ interval and in the high-$p_{T}$ region was around $12\%$. 
Neutral $\Xi$ was not measured by the STAR experiment. Thus, to calculate the feed-down from these particles, the yield of the neutral $\Xi$ was considered to be equal to the yield of the $\Xi^{\pm}$ particles.
The feed-down correction of the $\Lambda$ spectra from the $\Omega$ decay was not considered since it is negligible due to the lower yield of the $\Omega$ particles. The contribution from the $\Sigma_0$ decay to the $\Lambda$ yield was not subtracted as well due to the lack of $\Sigma^0$ yields measurements.

%==================================================
\subsection{Extrapolation of the $p_{T}$ spectra}
\label{subsec:Yields}
%==================================================

The final spectra for each particle species were obtained by dividing the raw yield $p_{T}$ distribution by the correction factors presented in the last section. 

As the spectra were not measured at low $p_T$ due to the limited acceptance coverage and at high $p_T$ due to the finite statistics, the spectra needed to be extrapolated in these two regions in order to determine the total production yield ($dN/dy$) as well as the mean transverse momentum ($\langle p_T \rangle$). 
This extrapolation was done with two different functional forms, an exponential function:
\begin{equation} \frac{d^2 N}{2 \pi p_T dp_T dy} \propto e^{-\frac{m_T}{T}}\end{equation}
and a Maxwell-Boltzmann function:
\begin{equation} \frac{d^2 N}{2 \pi p_T dp_T dy} \propto m_{T} e^{-\frac{m_{T}}{T}}\end{equation}

The spectra were better described with a Maxwell-Boltzmann function, resulting in fits with lower $\chi^2$ values than the exponential function fits.
In addition, the Maxwell-Boltzmann function was also used in the analysis of the data from Au+Au collisions at $200$~GeV and thus allows for a consistent comparison between the two data sets.

The main contribution for the determination of the $dN/dy$ comes from the low-$p_{T}$ region of the spectra. 
Therefore, in order to have a better estimate of the spectra shape in the low-$p_{T}$ region the fit was performed considering only the region of $p_{T}$ smaller than 1.5 GeV/c.
This extrapolation varied from 5 to 30\% of the final total yield for most particle species. 
In the case of the $\Omega$ particles, the extrapolated yield was on the order of 50\%, thus, the systematic uncertainty due to the extrapolation is large. 
The final $dN/dy$ was obtained by integrating the data in the measured region and using the fitted function in the low-$p_{T}$ region. 
The contribution of the high-$p_{T}$ region to the $dN/dy$ was verified and found to be negligible. 
The difference in the total integrated $dN/dy$ using the Maxwell-Boltzmann and the exponential function was considered to be the systematic uncertainty due to the extrapolation.

The $\langle p_{T} \rangle $ is obtained by integrating the whole range of the $p_{T}$ spectra as follows:
 
\begin{equation}\langle p_T \rangle = \frac{\int p_T \frac{dN}{dp_T} dp_T}{\int \frac{dN}{dp_T}dp_T}.\end{equation} 

The same functional forms used to determine the total yield were used to extrapolate the yield of the spectra in the unmeasured low-$p_{T}$ region to calculate the $\langle p_{T} \rangle $. 
The final uncertainty of the $\langle p_T \rangle $ values was estimated considering the variation of the point to point uncertainties of the measured particle spectra, and the uncertainty in the contribution from the extrapolated low-$p_{T}$ region.

%==================================================
\subsection{Systematic uncertainties due to the correction factors}
\label{subsec:Systemuncertainties}
%==================================================

Several sources of systematic uncertainties were evaluated and identified in the analysis. 
Variations in the detector performance and trigger efficiency during data taking and the dependence on position of the primary vertex were checked and found to be negligible. 
The main sources of the systematic uncertainties were the uncertainties in the correction factors of the measured spectra.

% Fig03
\begin{figure*}[hbt]
 \includegraphics[width=13cm]{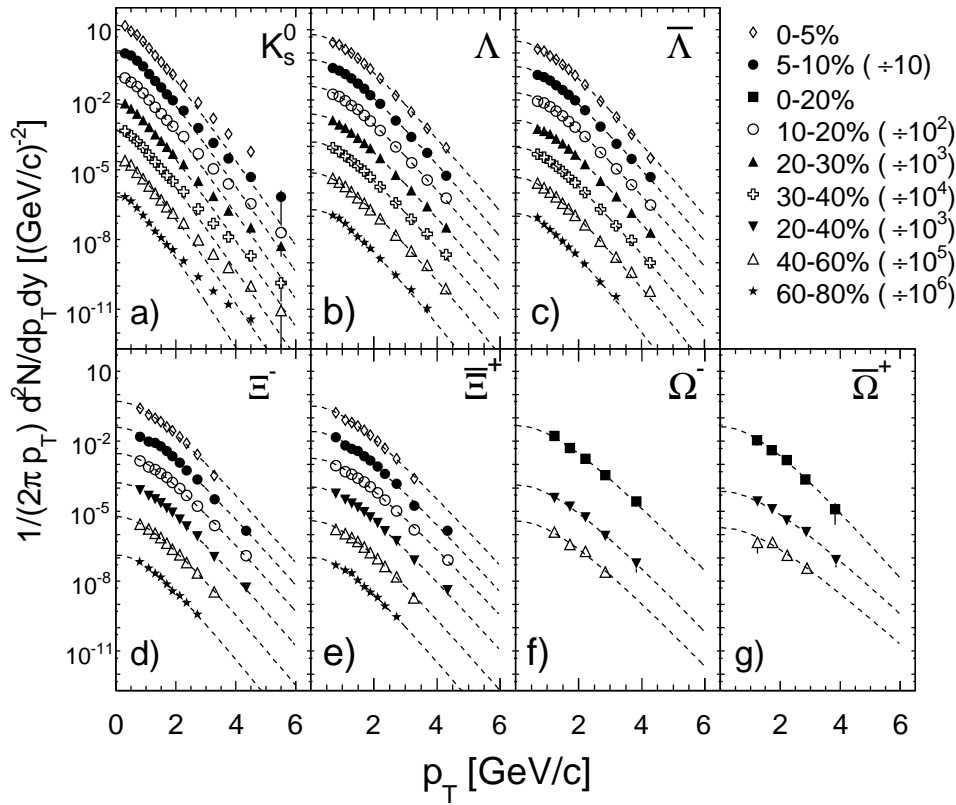} %
 \caption{Efficiency corrected $p_T$ spectra for the different centrality bins and for the various particles. Note that 7 centrality bins have been used for the $K^{0}_{S}$ and the $\Lambda$ while only 6 and 3 have been used for the $\Xi$ and $\Omega$, respectively. Errors are statistical only. The $\Lambda$ spectra are corrected for the feed-down of the $\Xi$ decay.}
\label{fig:AllSpectra}
\end{figure*}

In order to estimate the uncertainty due to the corrections of detector acceptance and efficiency of the analysis cuts, a careful comparison was performed between the distributions of the geometrical parameters obtained from the embedded simulated particles and the distributions of the tracks from real data.
A good agreement between the data and the simulated embedded distributions was obtained. 
The systematic uncertainties were calculated assuming a finite resolution to determine the cut value applied to the embedded sample during the correction factor calculation. 
The values of the cut were varied randomly given a certain assumed resolution and new correction factors were calculated. 
Next, the particle yields were obtained from these new correction factors. 
The variation of the yields obtained from this procedure was considered as one of the sources of the systematic uncertainty. 
This uncertainty was less than $10\%$ from all centrality bins. 

%==================================================
\section{Results and Discussions}
\label{sec:results}
%==================================================

The corrected $p_{T}$ spectra for K$^{0}_{S}$, $\Lambda$, $\Xi$, $\Omega$ and their anti-particles are presented in Fig.~\ref{fig:AllSpectra}. For better visualization, the spectra were divided by factors of 10, from central to peripheral data. 
The normalization factors are indicated in the figure.

% Fig04
\begin{figure}[hbt]
 \includegraphics[width=8.5cm]{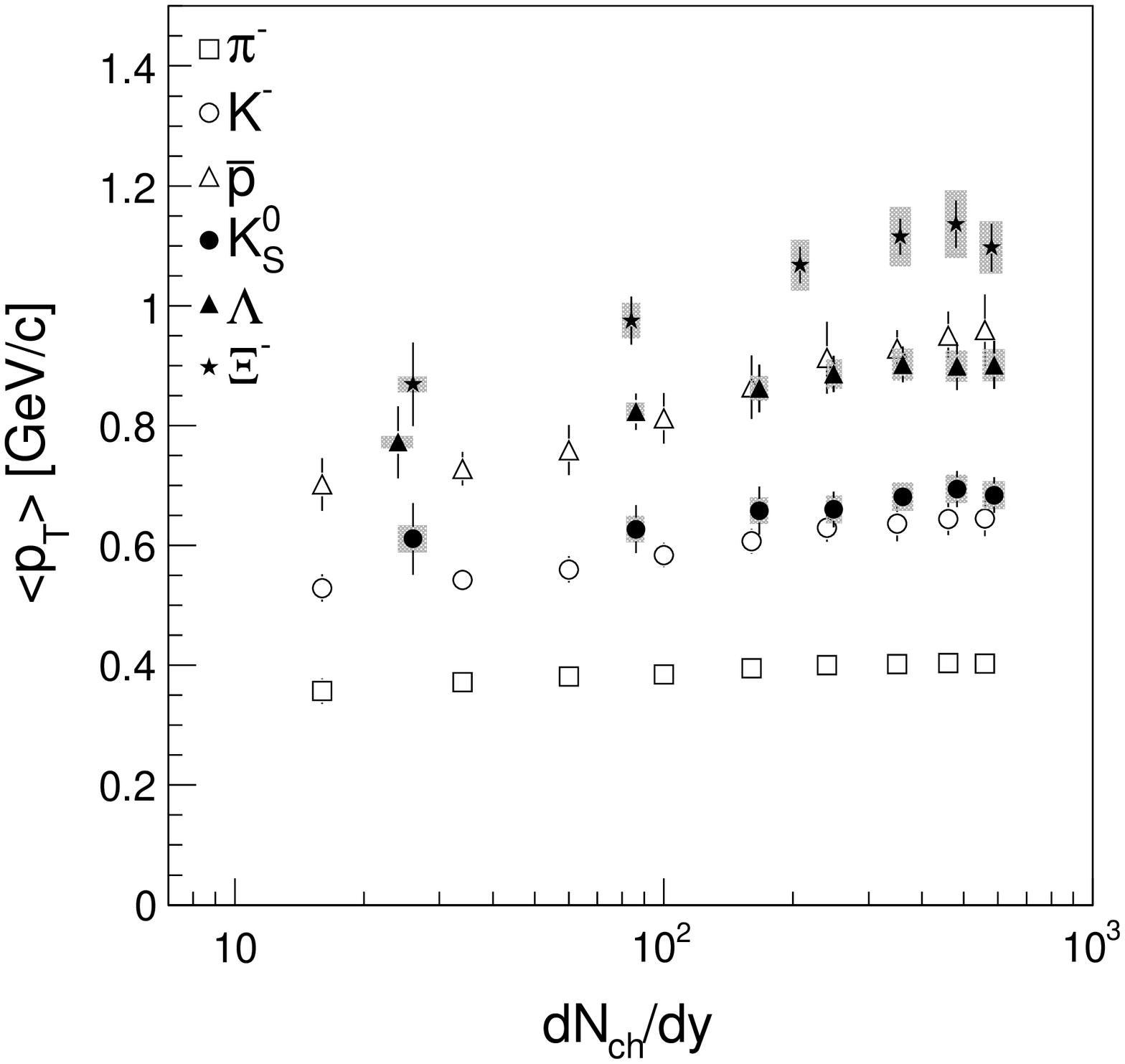}%
 \caption{Extrapolated average transverse momenta $\langle p_{T} \rangle$ as a function of $dN_{\rm ch}/dy$  for different particle species in Au+Au collisions at $62.4$~GeV. Statistical uncertainties are represented by the error bars at the points while the systematic uncertainties are represented by the gray bars. The $\pi$, charged $K$ and $p$ data were extracted from Ref. \cite{ref:Fuqiang}.}
 \label{fig:MeanPt}
\end{figure}

% Fig05
\begin{figure}[hbt]
 \includegraphics[width=8.5cm]{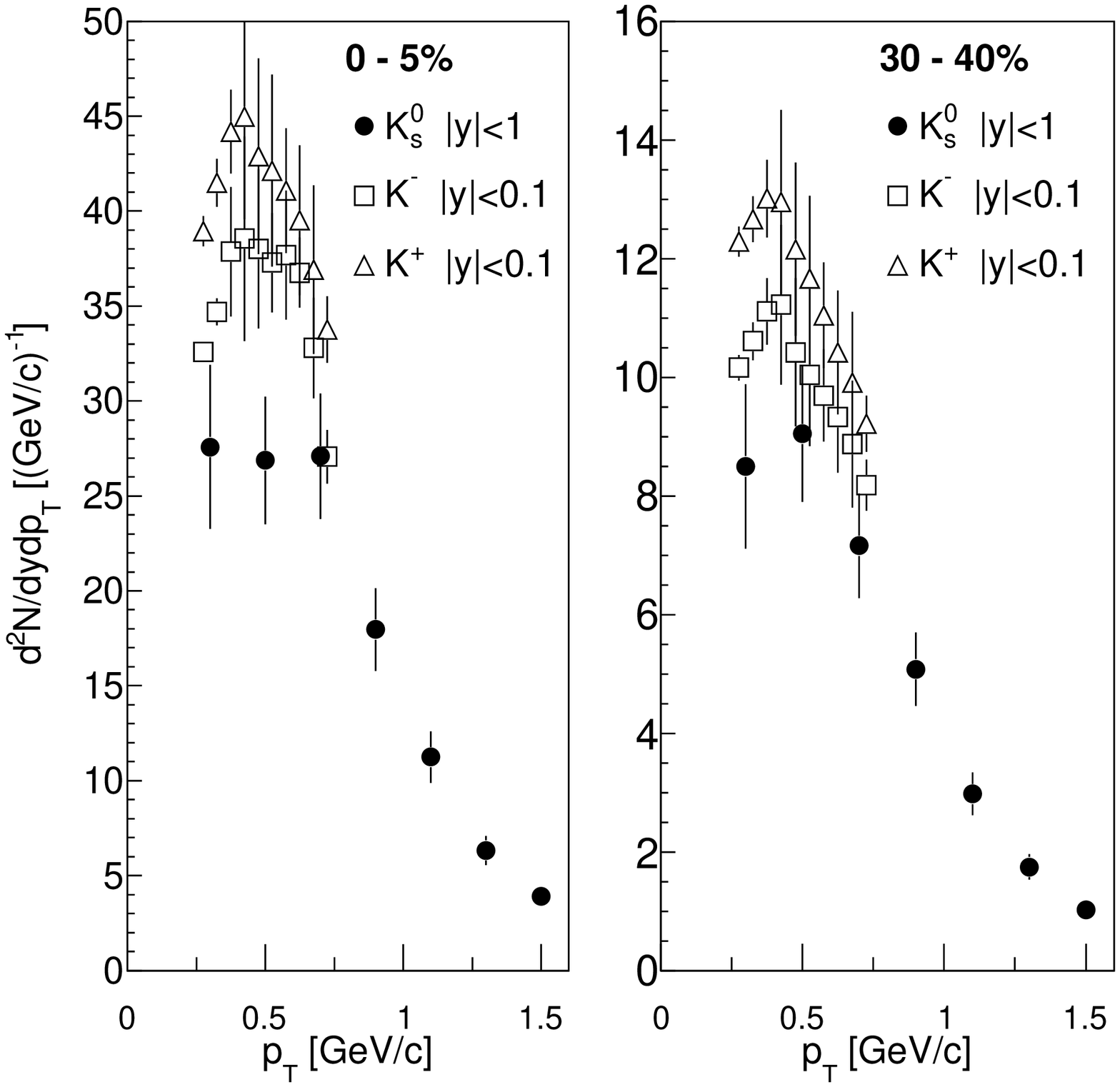}%
 \caption{$K^{0}_{S}$ $dN/dp_{T}$ spectra compared to the charged Kaon spectra for the event centrality of 0-5\% and 30-40\%.
 The charged Kaons data points are for rapidity range of $|y|<0.1$ and were extracted from Ref. \cite{ref:Fuqiang}.}
 \label{fig:KaonComp}
\end{figure}

The shape of the spectra in the low-$p_{T}$ region is sensitive to the characteristics of the evolution of the fireball such as transverse flow and the conditions of the kinetic freeze-out. With the increase in the momentum, the shape of the spectra is also affected by soft jets, whose fractional contribution to the shape of the spectra increases with increasing momentum.
The variations in the shape of the $p_{T}$ spectra for different particles and different centrality classes can be better visualized by examining the $\langle p_T \rangle $ values. 
The $\langle p_T \rangle $ calculated from the data, using a Maxwell-Boltzmann function to extrapolate for the unmeasured low-$p_{T}$ region, is presented in Table~\ref{tab:MeanPt}. 
The difference in the calculated $\langle p_T \rangle $ between using a Maxwell-Boltzmann or an exponential function is presented as the systematic uncertainty. 
Figure~\ref{fig:MeanPt} shows the evolution of $\langle p_T \rangle $ as a function of $dN_{\rm ch}/dy$ (see table \ref{tab:Centrality}) for different particle species. 

% Fig06
\begin{figure}
 \includegraphics[width=8.5cm]{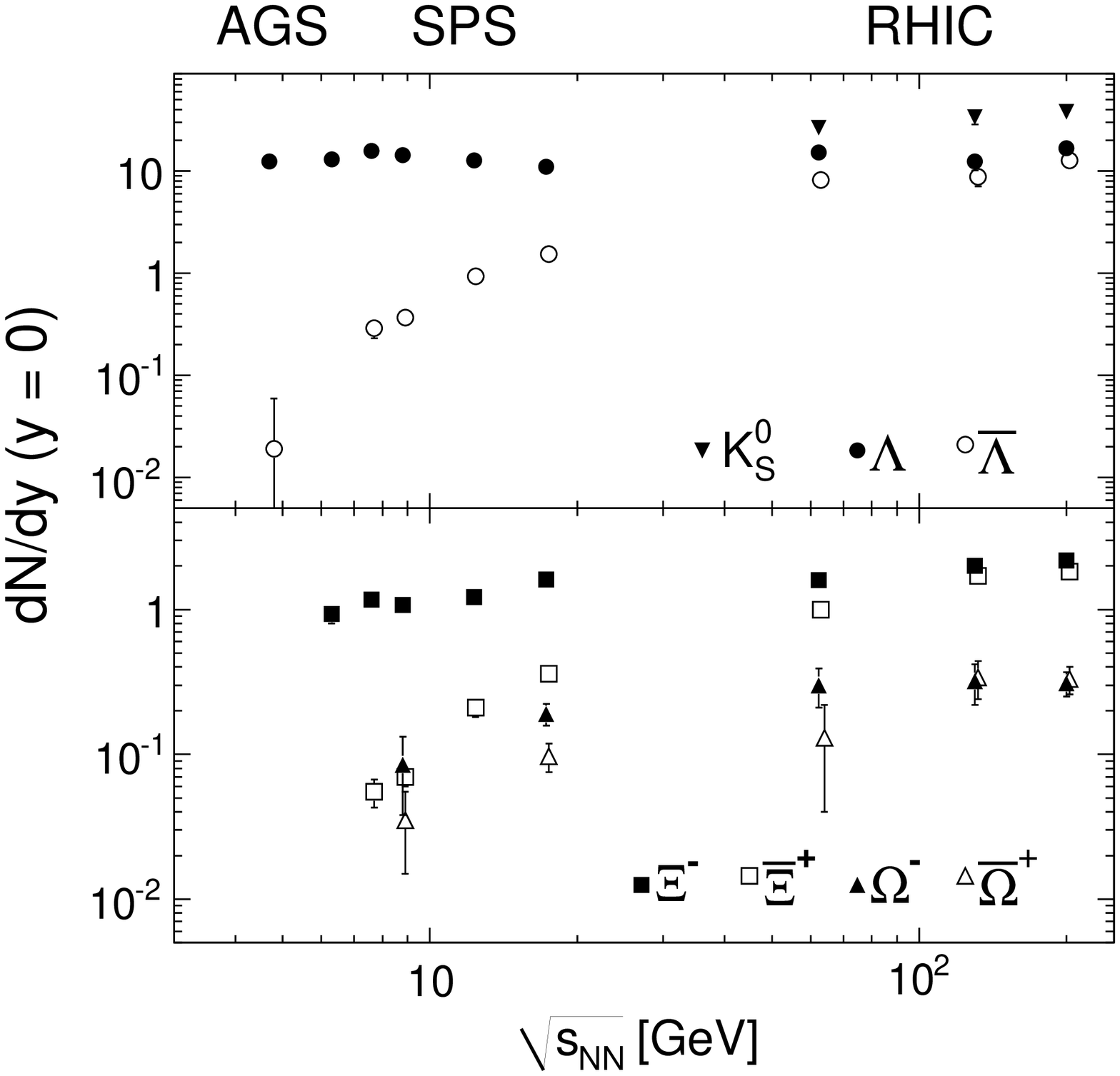}%
 \caption{Strange particle production yields at mid-rapidity in central Au+Au and Pb+Pb collisions versus the center of mass energy $\sqrt{s_{NN}}$. The top panel shows results for K$^0_S$ and $\Lambda$. The AGS values are from E896 \cite{ref:E896_Caines_JPG27_2001_311} (centrality $0-5$ \%). The SPS values are from NA49 \cite{ref:NA49_PRL93_022302_2004} (centrality $0-7$ \%) and the RHIC values are from STAR \cite{ref:STAR_PRL89_2002_092301, ref:STAR_PRL98_2007_062301} (centrality $0-5$ \%). For the multi-strange baryons $\Xi$ and $\Omega$ (bottom panel), the SPS results are from NA57 \cite{ref:NA57_PLB595_2004_68} (centrality $0-11$ \%) and the RHIC values are from STAR~\cite{ref:STAR_PRL92_2004_182301,ref:STAR_PRL98_2007_062301} (centrality $0-20$ \%).}
\label{fig:YieldSNN}
\end{figure}
 
The $\langle p_T \rangle $ values for $\bar{\Lambda}$ and $\bar{\Xi}^{+}$ are equal within uncertainties to the $\langle p_T \rangle $ for $\Lambda$ and $\Xi^{-}$, respectively, and thus they are not plotted in Fig.~\ref{fig:MeanPt}.
The $\langle p_T \rangle $ values from the $\Omega$ spectra are not included in this comparison due to the large uncertainties.
The $ \langle p_{T} \rangle $ for $\pi^{\pm}$, $K^{\pm}$ and $\bar{p}$ shown in Fig.~\ref{fig:MeanPt} were presented in Ref. \cite{ref:Fuqiang}. 
Error bars of each point correspond to the statistical uncertainty while the systematic uncertainty from the extrapolation is represented by the gray bands. 

%Tabel IV
\begin{table*}[hbt]
\begin{center}
\caption{Average transverse momenta, $\langle p_{T} \rangle$, in GeV/c, for the strange hadrons from Au+Au collisions at $62.4$~GeV. 
The first error is statistical, while the second is the systematic error arising from the extrapolation in the low $p_{T}$ region.}
\label{tab:MeanPt}
\begin{tabular}{c|c|c|c|c|c}
\hline\hline
 & \begin{minipage}[c][0.5cm][c]{0.33\columnwidth} K$^0_S$ \end{minipage} &\begin{minipage}[c][0.5cm][c]{0.33\columnwidth} $\Lambda$ \end{minipage} & \begin{minipage}[c][0.5cm][c]{0.33\columnwidth} $\overline{\Lambda}$ \end{minipage} & \begin{minipage}[c][0.5cm][c]{0.33\columnwidth} $\Xi^{-}$ \end{minipage} & \begin{minipage}[c][0.5cm][c]{0.33\columnwidth} $\bar{\Xi}^{+}$ \end{minipage} \\
\hline
\hline
 $0-5$\%   & 0.684 $\pm$ 0.030 $\pm$ 0.023 & 0.901 $\pm$ 0.041 $\pm$ 0.026 & 0.885 $\pm$ 0.043 $\pm$ 0.023 & 
1.097 $\pm$ 0.037 $\pm$ 0.043 & 1.130 $\pm$ 0.048 $\pm$ 0.050 
\\ \hline
 $5-10$\%  & 0.694 $\pm$ 0.027$\pm$ 0.023 & 0.899 $\pm$ 0.039 $\pm$ 0.026 & 0.892 $\pm$ 0.036 $\pm$ 0.025 & 
1.136 $\pm$ 0.040 $\pm$ 0.056 & 1.136 $\pm$ 0.048 $\pm$ 0.050
\\ \hline
 $10-20$\% & 0.681 $\pm$ 0.024 $\pm$ 0.023 & 0.902 $\pm$ 0.028 $\pm$ 0.026 & 0.887 $\pm$ 0.026 $\pm$ 0.025 & 
1.115 $\pm$ 0.028 $\pm$ 0.049 & 1.130 $\pm$ 0.037 $\pm$ 0.052
\\ \hline
 $20-30$\% & 0.660 $\pm$ 0.034 $\pm$ 0.023 & 0.886 $\pm$ 0.032 $\pm$ 0.024 & 0.892 $\pm$ 0.025 $\pm$ 0.026 & 
\multirow{2}{*} {1.068 $\pm$ 0.025 $\pm$ 0.042 }& \multirow{2}{*} {1.042 $\pm$ 0.031 $\pm$ 0.035 } 
\\ \cline{1-4}
 $30-40$\% & 0.658 $\pm$ 0.036 $\pm$ 0.022 & 0.862 $\pm$ 0.035 $\pm$ 0.021 & 0.849 $\pm$ 0.029 $\pm$ 0.020 & 
\\
\hline
 $40-60$\% & 0.627 $\pm$ 0.038 $\pm$ 0.022 & 0.823 $\pm$ 0.035 $\pm$ 0.015 & 0.807 $\pm$ 0.032 $\pm$ 0.013 & 
0.975 $\pm$ 0.039 $\pm$ 0.029 & 1.022 $\pm$ 0.047 $\pm$ 0.041 \\
\hline
 $60-80$\% & 0.611 $\pm$ 0.062 $\pm$ 0.022 & 0.772 $\pm$ 0.058 $\pm$ 0.010 & 0.766 $\pm$ 0.075 $\pm$ 0.010 & 
0.869 $\pm$ 0.074 $\pm$ 0.013 & 0.90 $\pm$ 0.11 $\pm$ 0.02 \\

\hline
\hline
\end{tabular}
\end{center}
\end{table*}

The $\langle p_T \rangle $ values of $K^{0}_{S}$ are in agreement with the values of $K^{-}$ and the $\Lambda$ values with those of $\bar{p}$ spectra.
The $\langle p_T \rangle $ shows a trend of increase from peripheral to central collisions for all particles. 
This increase is more pronounced for the heavier particles ($\bar{p}$) than for the lighter particles ($\pi$). 
These observations are consistent with a collective radial flow among formed hadrons, which increases with centrality.

%Table V
\begin{table*}[hbt]
\begin{center}
\caption{Integrated yield, $dN/dy$, for $K^{0}_{S}$, $\Lambda$ and $\overline{\Lambda}$ measured in Au+Au collisions at $62.4$~GeV using data and a Maxwell-Boltzmann function for the extrapolation to the unmeasured low $p_{T}$ region. Quoted uncertainties are the statistical errors and the systematic uncertainties. The $\Lambda$ and $\overline{\Lambda}$ yields are corrected by subtracting the contribution of the feed-down from the $\Xi$ weak decays.}
\label{tab:Yield}
\begin{tabular}{c|c|c|c}
\hline\hline
 & \begin{minipage}[c][0.5cm][c]{0.33\columnwidth} K$^0_S$  \end{minipage} & \begin{minipage}[c][0.5cm][c]{0.33\columnwidth} $\Lambda$ \end{minipage} & \begin{minipage}[c][0.5cm][c]{0.33\columnwidth} $\overline{\Lambda}$  \end{minipage} \\
\hline
\hline
\quad $0-5$\%   \quad&\quad 27.4 $\pm$ 0.6 $\pm$ 2.9 \quad&\quad 15.7 $\pm$ 0.3 $\pm$ 2.3     \quad&\quad 8.3 $\pm$ 0.2 $\pm$ 1.1 
\\ \hline
\quad $5-10$\%  \quad&\quad 21.9 $\pm$ 0.5 $\pm$ 2.3 \quad&\quad 12.2 $\pm$ 0.3 $\pm$ 1.9     \quad&\quad 6.1 $\pm$ 0.1 $\pm$ 0.8  
\\ \hline
\quad $10-20$\% \quad&\quad 17.1 $\pm$ 0.3 $\pm$ 1.7 \quad&\quad 9.1 $\pm$ 0.2 $\pm$ 1.3   \quad&\quad 4.7 $\pm$ 0.1 $\pm$ 0.6 
\\ \hline
\quad $20-30$\% \quad&\quad 12.1 $\pm$ 0.3 $\pm$ 1.1 \quad&\quad 6.2 $\pm$ 0.1 $\pm$ 0.8   \quad&\quad 2.99 $\pm$ 0.05 $\pm$ 0.40 
\\ \hline
\quad $30-40$\% \quad&\quad 8.1 $\pm$ 0.2 $\pm$ 0.7 \quad&\quad 4.1 $\pm$ 0.1 $\pm$ 0.6 \quad&\quad 2.25 $\pm$ 0.04 $\pm$ 0.30 
\\ \hline
\quad $40-60$\% \quad&\quad 4.0 $\pm$ 0.1 $\pm$ 0.3 \quad&\quad 2.01 $\pm$ 0.04 $\pm$ 0.26 \quad&\quad 1.16 $\pm$ 0.02 $\pm$ 0.16 
\\ \hline
\quad $60-80$\% \quad&\quad 1.13 $\pm$ 0.05 $\pm$ 0.09 \quad&\quad 0.504 $\pm$ 0.017 $\pm$ 0.07 \quad&\quad 0.343 $\pm$ 0.012 $\pm$ 0.036 
\\ \hline
\hline
\end{tabular}
\end{center}
\end{table*}

The integrated yields for each centrality selection for the $K^{0}_{S}$, $\Lambda$ and $\bar{\Lambda}$ are presented in Table~\ref{tab:Yield} while the yields for the $\Xi^{-}$, $\bar{\Xi}^{+}$, $\Omega^{-} $ and $\bar{\Omega}^{+}$  are presented in Table~\ref{tab:Yield2}. 
The systematic uncertainties correspond to the uncertainties due to the correction factors and the uncertainty due to the extrapolation.

STAR experiment has presented $K^{\pm}$ yields and spectra for the same colliding system in Ref.~\cite{ref:Fuqiang}.
For the most central events, the integrated yields of $K^{0}_{S}$ is lower than the average between the integrated yields of $K^{+}$ and $K^{-}$.
Figure~\ref{fig:KaonComp} shows the comparison of the $K^{0}_{S}$ and $K^{\pm}$ $d^2N/dydp_{T}$ spectra in both top 5\% and 30-50\% centrality bins. 
The data points are consistent within uncertainties for the $p_{T}$ range of 0.6-0.8~GeV/c but the $K^{0}_{S}$ data points are significantly lower for the two lowest $p_{T}$ bins. 
The difference is no larger than two sigmas.
The charged Kaon spectra is limited by the dE/dx identification method to a maximum $p_{T}$ of 0.75~GeV/c.
In addition, the charged Kaon yields presented in Ref.~\cite{ref:Fuqiang} are measured in the rapidity interval of $|y|<0.1$ whereas the $K^{0}_{S}$ yields presented are for rapidity interval of $|y|<1.0$.
We also note that the difference between the charged Kaon spectra and the $K^{0}_{S}$ spectra decrease with decreasing centrality and in the peripheral bin of 30-40\% the data points are in agreement within the experimental error bars for all $p_{T}$ points in the overlap region.

% Fig07
\begin{figure}
 \includegraphics[width=8.5cm]{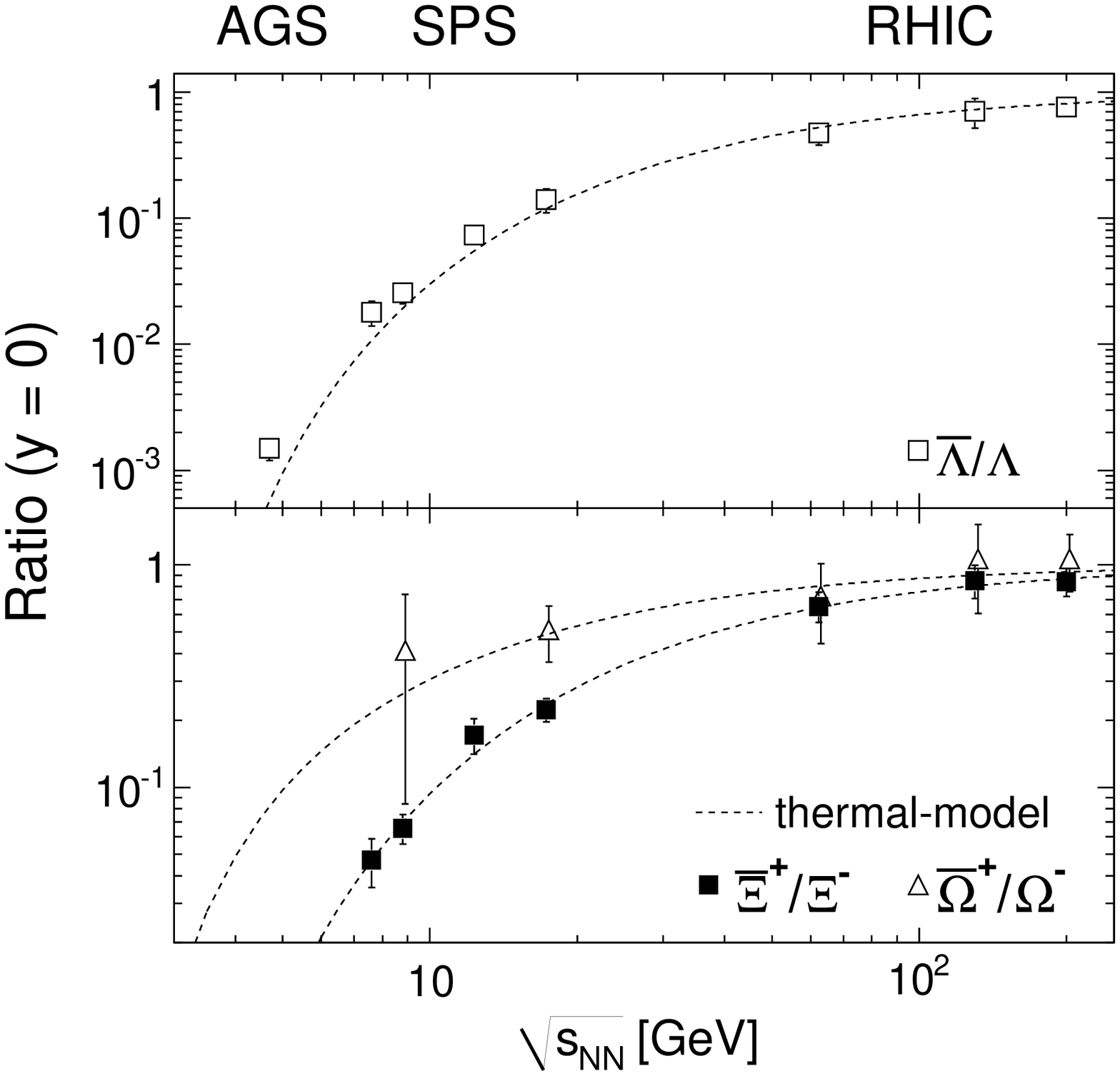}%
 \caption{Anti-baryon to baryon yield ratios for strange baryons versus the center of mass energy $\sqrt{s_{NN}}$. $\overline{\Lambda}/\Lambda$ is shown in the top panel while the multi-strange baryons are on the bottom panel. The data from AGS are not corrected for the weak decay feed-down from the multistrange baryons while the data from SPS and RHIC are corrected. The lines are the results of a thermal model calculation (see text section \ref{sec:Chemical}). The AGS values are from E896 \cite{ref:E896_Caines_JPG27_2001_311} (centrality $0-5$ \%). The SPS values are from NA49 \cite{ref:NA49_PRL93_022302_2004} (centrality $0-7$ \%) and the RHIC values are from STAR \cite{ref:STAR_PRL89_2002_092301, ref:STAR_PRL98_2007_062301} (centrality $0-5$ \%). For the multi-strange baryons $\Xi$ and $\Omega$ (bottom panel), the SPS results are from NA57 \cite{ref:NA57_PLB595_2004_68} (centrality $0-11$ \%) and the RHIC values are from STAR~\cite{ref:STAR_PRL92_2004_182301,ref:STAR_PRL98_2007_062301} (centrality $0-20$ \%).}
\label{fig:RatioSNN}
\end{figure}
  
% Fig08
\begin{figure}[hbt]
 \includegraphics[width=8.5cm]{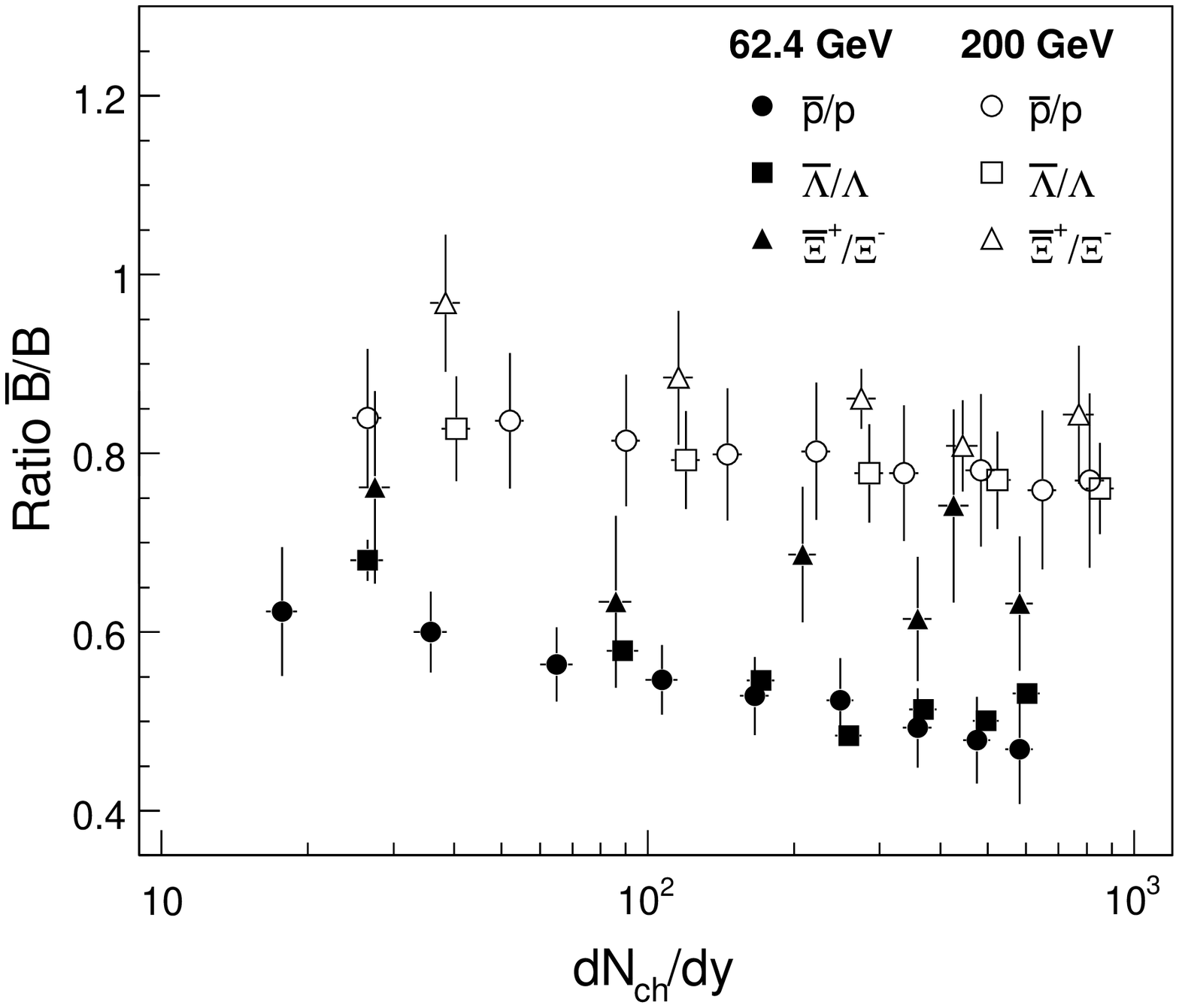}%
 \caption{Anti-baryon to baryon yield ratios for strange particles and protons as a function of $dN_{\rm ch}/dy$ at $\sqrt{s_{NN}}=62.4$ and $200$~GeV. The $p$ data were extracted from Ref. \cite{ref:Fuqiang}. The $\sqrt{s_{NN}}=200$~GeV strange hadron data were extracted from \cite{ref:STAR_PRL98_2007_062301}.}
 \label{fig:RatiovsCentr}
\end{figure}

Figure~\ref{fig:YieldSNN} shows the dependence of the singly-strange and multi-strange particle yields on the collision energy,
including the results from this analysis. 
The lowest energy points are from Au+Au collisions measured by the E896 experiment at the AGS~\cite{ref:E896_Caines_JPG27_2001_311} and the points in the SPS energy region are from Pb+Pb collisions measured by the NA49 experiment~\cite{ref:NA49_PRL93_022302_2004} and the NA57 experiment \cite{ref:NA57_PLB595_2004_68} . 
The RHIC data points for 130 GeV and $200$~GeV~\cite{ref:STAR_PRL92_2004_182301,ref:STAR_PRL98_2007_062301} are also presented. 
The strange baryon and meson production exhibits only a small energy dependence from AGS through SPS and on up to the top RHIC energies. 
From the AGS energy to the top SPS energy region, the $\Lambda$ yield seems to increase at first, reaching a maximum around $\sqrt{s_{NN}} \approx 10$~GeV and then starts to decrease slightly with energy. 
At RHIC the yield of $\Lambda$ seems to stay approximately constant, independent of the collision energy. 

Meanwhile, the $\Omega$ and the anti-baryon yields show a clear and smooth rise with energy.
The overall trend of the strange and multi-strange anti-particle yields with energy indicates a transition from a transport dominated production mechanism at the lower energies (AGS, SPS) to a production dominated by quark pair creation at the top RHIC energy. Transport dominated means that the initial baryon number of the colliding nuclei is transported to the mid-rapidity region.
This transition occurs in a way that the mid-rapidity baryon-free regime is reached smoothly. 

Figure~\ref{fig:RatioSNN} shows the anti-baryon to baryon ratio of $\Lambda$, $\Xi$ and $\Omega$ baryons as a function of energy which can be seen to reach almost unity for all particle species at $\sqrt{s_{NN}}=200$~GeV. 
The dashed lines in Fig.~\ref{fig:RatioSNN} show the results of a thermal model calculation \cite{ref:Cleymans_all} that will be  discussed in section \ref{sec:Chemical}. 

% Table VI
\begin{table*}[hbt]
\begin{center}
\caption{Integrated mid-rapidity yield, $dN/dy$, for $\Xi^{-}$, $\bar{\Xi}^{+}$, $\Omega^{-}$ and $\bar{\Omega}^{+}$ measured in Au+Au collisions at $62.4$~GeV (using a Maxwell-Boltzmann function for the extrapolation). Quoted uncertainties are the statistical errors and the systematic uncertainties.}
\label{tab:Yield2}
\begin{tabular}[b]{c|c|c|c|c}
\hline\hline
\quad & \begin{minipage}[c][0.5cm][c]{0.33\columnwidth} $\Xi^{-}$  \end{minipage} & \begin{minipage}[c][0.5cm][c]{0.33\columnwidth} $\bar{\Xi}^{+}$  \end{minipage} & \begin{minipage}[c][0.5cm][c]{0.33\columnwidth} $\Omega^{-} $  \end{minipage} & \begin{minipage}[c][0.5cm][c]{0.33\columnwidth} $\bar{\Omega}^{+}$ \end{minipage}\\
\hline
\hline
\quad $0-5$\%   \quad&\quad 1.63 $\pm$ 0.09 $\pm$ 0.18 \quad&\quad 1.03 $\pm$ 0.09 $\pm$ 0.11 \quad&\quad \multirow{3}{*}{0.212 $\pm$ 0.028 $\pm$ 0.018} \quad&\quad \multirow{3}{*}{0.167 $\pm$ 0.027 $\pm$ 0.015}
\\ \cline{1-3}
\quad $5-10$\%  \quad&\quad 1.16 $\pm$ 0.06 $\pm$ 0.16 \quad&\quad 0.86 $\pm$ 0.08 $\pm$ 0.12 \quad&\quad \quad&\quad  
\\ \cline{1-3}
\quad $10-20$\% \quad&\quad 0.96 $\pm$ 0.04 $\pm$ 0.11 \quad&\quad 0.59 $\pm$ 0.03 $\pm$ 0.06 \quad&\quad \quad&\quad  
\\ \hline
\quad $20-40$\% \quad&\quad 0.52 $\pm$ 0.02 $\pm$ 0.06 \quad&\quad 0.357 $\pm$ 0.016 $\pm$ 0.037 \quad&\quad 0.056 $\pm$ 0.006 $\pm$ 0.005 \quad&\quad 
0.038 $\pm$ 0.005 $\pm$ 0.003 
\\ \hline
\quad $40-60$\% \quad&\quad 0.183 $\pm$ 0.008 $\pm$ 0.021 \quad&\quad 0.116 $\pm$ 0.005 $\pm$ 0.017 \quad&\quad 0.0165 $\pm$ 0.0027 $\pm$ 0.0014 \quad&\quad 
0.0103 $\pm$ 0.0020 $\pm$ 0.0010 
\\ \hline
\quad $60-80$\% \quad&\quad 0.042 $\pm$ 0.003 $\pm$ 0.005 \quad&\quad 0.032 $\pm$ 0.003 $\pm$ 0.004 \quad&\quad  \quad&\quad 
\\ 
\hline \hline
\end{tabular}
\end{center}
\end{table*}

Figure~\ref{fig:RatiovsCentr} shows the anti-baryon to baryon ratio of $p$, $\Lambda$ and $\Xi$ baryons as a function of the centrality of the collisions, represented by the measured number of charged hadrons per unit of rapidity ($dN_{\rm ch}/dy$). Results from $\sqrt{s_{NN}}=200$~GeV collisions (open symbols) and from the $\sqrt{s_{NN}}=62.4$~GeV collisions (solid symbols) are compared.
At $\sqrt{s_{NN}}=62.4$~GeV, the $\bar{p} / p$ and $\bar{\Lambda} / \Lambda$ ratios show a similar trend, decreasing with increasing centrality, consistent with a higher degree of baryon number transport 
 from beam rapidity for central collisions. 
The decrease with centrality seems to be larger at $62.4$~GeV than at $200$~GeV, which is consistent with a higher fraction of net-baryons being transported from beam rapidity to mid-rapidity at the lower collision energies. 
%The $\Xi$ is formed from two strange quarks and one light quark, and should be less sensitive to the net-baryon density and consequently show smaller variation with centrality.
At $\sqrt{s_{NN}}=200$~GeV where there is a wider rapidity gap between the beam particles, the net-baryon density at mid-rapidity is lower, thus, the anti-baryon to baryon ratios of the studied particles are higher than at $62.4$~GeV. In addition, the ratio is independent of centrality at $\sqrt{s_{NN}}=200$~GeV, indicating that pair production is dominating and that there is a smaller variation of net-baryon density with centrality.

%============================================================
\subsection{Chemical properties} 
\label{sec:Chemical}
%============================================================

The chemical properties of the bulk particle production can be addressed by using thermal-statistical models \cite{ref:Becattini,ref:Cleymans,ref:PBM,ref:Rafelski,ref:Kaneta_Xu}. They are applicable under the assumption that the yields of particles are governed by statistical laws. The particle abundance of species $i$ per system volume $(N_{i}/V)$ can be parameterized by:
\begin{equation} \frac{N_{i}}{V} = \frac{g_{i}}{(2\pi)^3} \int{\left[ \gamma_{S}^{-\left|S'_{i}\right|} \exp(\frac{E_{i}-\mu_{i}}{T_{\rm ch}}) \pm 1 \right]^{-1} d^{3}p}\end{equation}
where
\begin{equation} \mu_{i} = \mu_{B}B_{i}-\mu_{Q}Q_{i}-\mu_{S}S_{i} \end{equation}
and the $g_{i}$ factor is the spin-isospin degeneracy, $T_{\rm ch}$ is the chemical freeze-out temperature that marks the end of inelastic interactions, $\mu_{Q}$, $\mu_{B}$, $\mu_{S}$ are the charge, baryon and strangeness chemical potentials respectively, $B_{i}$, $Q_{i}$ and $S_{i}$ are the baryon, charge and strangeness numbers, $E_{i}$ is the energy of the particle, $\gamma_s$ is the strange quark phase-space occupancy and $S'_{i}$ is the number of valence strange and anti-strange quarks in the particle $i$. 
 
% Fig09
\begin{figure}
 \includegraphics[width=8.5cm]{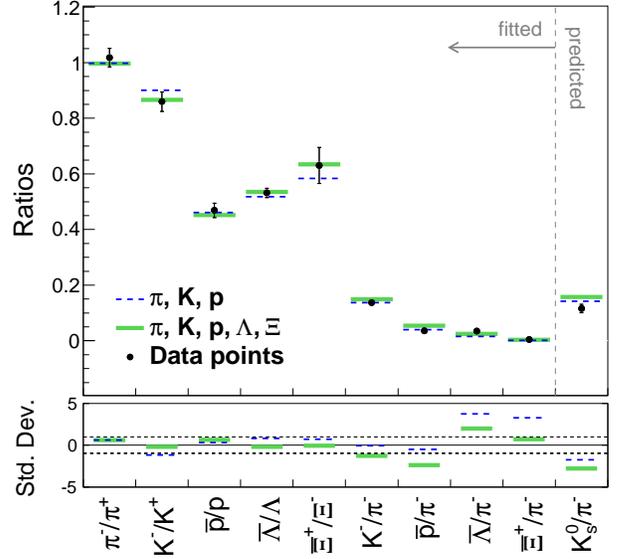}%
 \caption{(color online) Particle yield ratios as obtained by measurements (black dots) for the most central ($0-5$\%) Au+Au collisions at $62.4$~GeV, and statistical model predictions (lines). 
 The ratios indicated by the dashed lines (blue) were obtained by using only $\pi$, K and protons, while the ratios indicated by the full lines (green) were obtained by also using the hyperons in the fit.}
\label{fig:Stat_Ratios}
\end{figure}

These thermal-statistical models have been successful in reproducing the particle ratios for a large range of energies and also have been able to reproduce ratios including strange and multi-strange particles such as the $\Omega/\pi$ ratio~\cite{ref:Jun_1,ref:PBM_1999}. 
The evolution of the anti-baryon to baryon ratio of the studied particles with $\sqrt{s_{NN}}$ is also well reproduced in these models as indicated in Fig.~\ref{fig:RatioSNN} by the dashed lines. 
These lines correspond to the anti-baryon to baryon ratios using a thermal model calculation with parameters (temperature and chemical potentials) given by  the systematic study presented in Ref.~\cite{ref:Cleymans06}. 
The data at the intermediate $62.4$~GeV energy are also well reproduced.

In order to perform a thermal fit to the STAR data, we used the model THERMUS described in \cite{ref:THERMUS}. 
The parameters considered in this thermal fit are the temperature ($T_{\rm ch}$), the baryon chemical potential ($\mu_B$) 
and the strangeness saturation factor ($\gamma_s$).
The strangeness chemical potential ($\mu_s$) was constrained by the initial $S/V$ ratio (strangeness per volume).
The charge chemical potential ($\mu_Q$) was fixed to zero and the system volume was not considered because particle ratios were used.
To calculate the particle ratios, proton yields were corrected to subtract the contributions from the weak decays of $\Lambda$ particles.
The feed-down contribution from the $\Sigma$ decay into the proton and anti-proton yields were done within the thermal model and the correction was on the order of 7\%.
The pions were corrected for the weak decay feed-down, muon contamination and
background pions produced in the detector materials. The $\Lambda$ yields were corrected for the weak decay feed-down from the $\Xi$.
Figure~\ref{fig:Stat_Ratios} shows the particle ratios as obtained by measurements for the most central data as black dots. 

% Fig10
\begin{figure}
 \includegraphics[width=8.5cm]{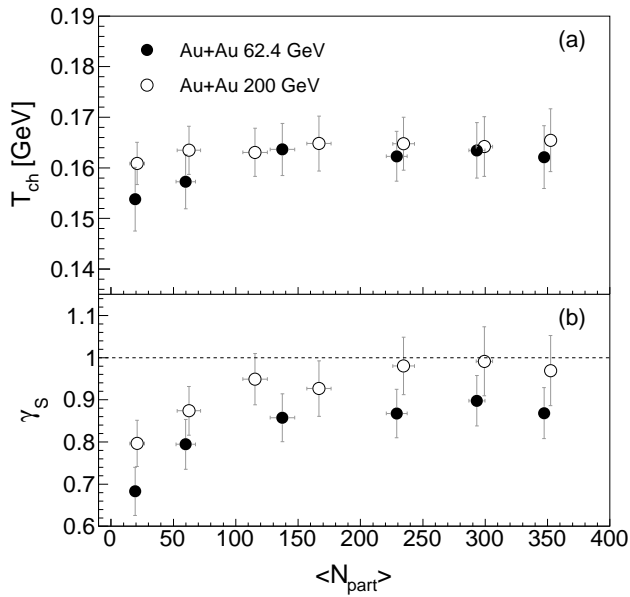}
 \caption{Chemical freeze-out temperature $T_{\rm ch}$ (panel a) and strangeness saturation factor $\gamma_s$ (panel b) as a function of the mean number of participants.}
\label{fig:Stat_GammaS}
\end{figure}

The fit procedure was performed with two different data sets, considering first only the particle ratios of $p$, $\bar{p}$, $\pi^{\pm}$ and $K^{\pm}$. 
Results from this fit are shown in Fig.~\ref{fig:Stat_Ratios} as dashed lines (blue online) and it is seen that the model fits quite well the ratios of pions, protons and charged kaons, but it under predicts the ratio $\bar{\Xi}^{+} / \Xi^{-}$. 
The thermal fit was repeated considering also the particle ratios with $\Lambda$, $\bar{\Lambda}$ and $\Xi^{\pm}$. 
The $\Omega$ ratios were not included in these thermal fits because the yields were obtained for different event centrality bins due to limited statistics in data. 
The results of the thermal fit, including the $\Lambda$ and $\Xi$ ratio are also shown in Fig.~\ref{fig:Stat_Ratios} by the full lines (green online).
The standard deviations of the data to the thermal fit result are also shown at the bottom of Fig.~\ref{fig:Stat_Ratios}. 
Most of the results are within one standard deviation from the data for the case where the strange particle ratios are also considered in the calculation. 
The inclusion of the strange baryons in the thermal studies results in a more constrained fit and a better description of these yields. 
In addition there is a small increase of $T_{\rm ch}$ from (147 $\pm$ 7)~MeV to (163 $\pm$ 5)~MeV but the value of $\gamma_s$ remains constant at 0.88~$\pm$~0.06. 
The particle ratios including the strange baryons are well described by the thermal fit, but it over-predicts the production of $\bar{p}$ with respect to pions. 

Figure~\ref{fig:Stat_GammaS} shows the evolution of the chemical freeze-out temperature (top panel) and the strangeness phase-space occupancy factor (bottom) with the number of participants. 
The solid symbols represent the results for $\sqrt{s_{NN}}=62.4$~GeV and the open symbols represent the result for $\sqrt{s_{NN}}=200$~GeV. 
$T_{\rm ch}$ is independent of the centrality of the collision and its value is consistent with the top RHIC energy measurements. 
In contrast, $\gamma_{s}$ increases from a lower value for peripheral collisions to saturate at a value consistent with unity for the most central events at $\sqrt{s_{NN}}=200$~GeV and close to unity (0.87~$\pm$~0.07) at $\sqrt{s_{NN}}=62.4$~GeV. 
These results are similar to the results obtained previously at the higher RHIC energies, where $\gamma_{s}$ rises from peripheral collisions and reaches values consistent with unity for the most central events~\cite{ref:Jun02}. 

% Fig11
\begin{figure}
 \includegraphics[width=8.5cm]{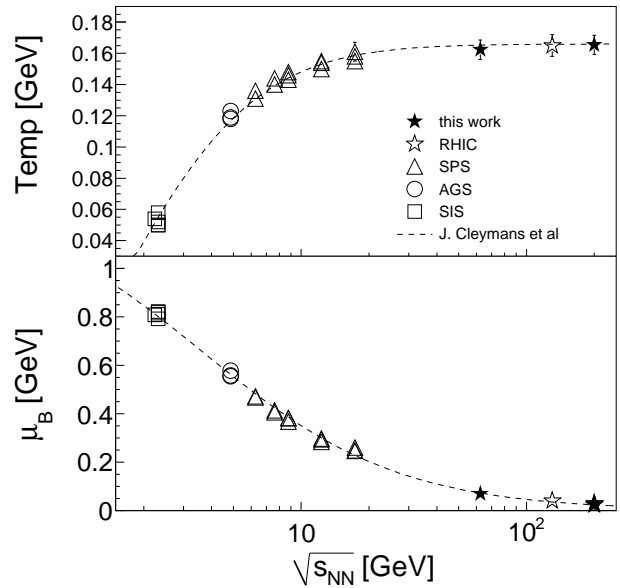}
 \caption{Temperature and baryon chemical potential obtained from thermal model fits as a function of $\sqrt{s_{NN}}$ (see \cite{ref:Cleymans_all} ). The dashed lines correspond to the parameterizations given in \cite{ref:Cleymans_all}. The solid stars show the result for $\sqrt{s_{NN}}=62.4$ and $200$~GeV.}
 \label{fig:TvsSnn}
\end{figure}
 
Figure~\ref{fig:TvsSnn} shows the energy dependence of the chemical freeze-out temperature and the baryon chemical potential. 
The open symbols were extracted from~\cite{ref:Cleymans_all}. The solid stars show the values obtained from this analysis for $\sqrt{s_{NN}}=62.4$ and for the $200$~GeV STAR data with the same particle ratios. These results are consistent with other thermal fits presented in Ref.~\cite{ref:Cleymans_all} for the same energy range, where data points from other RHIC experiments are considered.
The dashed curves in Fig.~\ref{fig:TvsSnn} are the numerical parameterizations of $T_{\rm ch}$ and $\mu_{B}$ presented in Ref.~\cite{ref:Cleymans06} using:
\begin{equation} \mu_{B}(\sqrt{s}) = \frac{1.308 \rm \,GeV}{1+0.273 \rm \,GeV^{-1} \sqrt{\it{s}}} \end{equation}
and
\begin{equation} T_{\rm ch}=0.166 \rm \,GeV - 0.139 \rm \,GeV^{-2} \mu_{B}^{2}-0.053 \rm \,GeV^{-4} \mu_{B}^{4}\end{equation}

The values obtained in this analysis for $T_{\rm ch}$ and $\mu_{\rm B}$ from the Au+Au $62.4$~GeV data are in good agreement with the parameterizations that were obtained fitting the data points at higher and lower energies. 
As the collision energy increases, the baryon chemical potential decreases continuously reaching very small values at RHIC. 
In contrast, the chemical freeze-out temperature $(T_{\rm ch})$ seems to achieve saturation at approximately 160 MeV and all three RHIC energy points shown in Fig.~\ref{fig:TvsSnn} are already at this temperature.

%============================================================
\subsection{Strangeness Enhancement}
\label{sec:enhance}
%============================================================

Strangeness enhancement is traditionally defined as the ratio of the strange particle yield measured in heavy-ion collisions, normalized by the mean number of participant nucleons ($\langle N_{part} \rangle$) and the yield measured in $p+p$ collisions.
Strangeness enhancement was observed at the lower SPS energies~\cite{ref:NA57} as well as at higher RHIC energies~\cite{ref:HelenSTAR}.
The multi-strange baryons show a higher enhancement factor than the $\Lambda$ particles, which is consistent with the picture of the enhancement
of $s\bar{s}$ pair production in a dense partonic medium as opposed to a pure hadronic medium \cite{ref:Rafelski_2}. 
It is also argued that part of the enhancement can be due to canonical suppression effects in $p+p$ collisions~\cite{ref:Redlich}. In this scenario, the production of strange hadrons is actually suppressed in $p+p$ collisions due to limitations of phase space.
To discriminate among the different mechanisms that can affect the strangeness enhancement factors, it is
necessary to study the systematic behavior for different collision energies, systems and centralities.

% Fig12
\begin{figure}[hbt]
 \includegraphics[width=8.5cm]{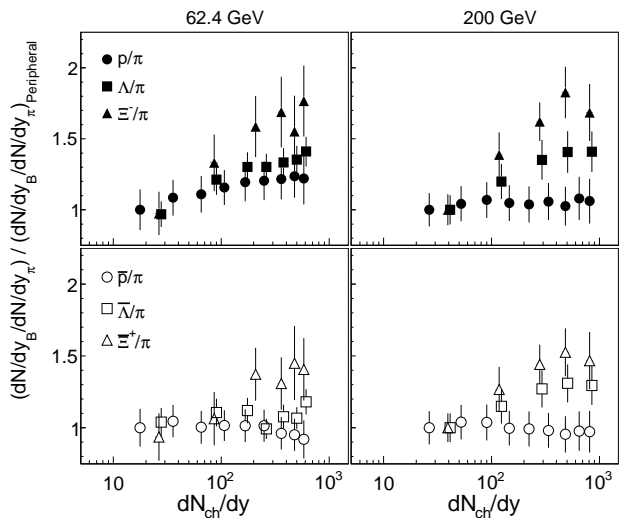} 
 \caption{Ratio of baryon (solid symbols) and anti-baryon (open symbols) to $\pi^{+}$ as a function of $dN_{\rm ch}/dy$ for $\sqrt{s_{NN}}=62.4$~GeV (left panel) and $\sqrt{s_{NN}}=200$~GeV (right panel). The $\pi$ and $p$ data were extracted from Ref. \cite{ref:Fuqiang}.}
 \label{fig:BarionvsCentr}
\end{figure}

Since there are no data on strange baryon production for $p+p$ collisions at $\sqrt{s_{NN}}=62.4$~GeV for comparison, we show the ratios of strange baryons and $\pi$ mesons in Au+Au collisions as a function of $dN_{\rm ch}/dy$ normalized to the ratio in the most peripheral centrality bin (Fig.~\ref{fig:BarionvsCentr}). 
The $\Omega$ data points were not included in this plot due to the difference in the centrality selections.
The measurements at $\sqrt{s_{NN}}=62.4$~GeV (left panels) and $200$~GeV (right panels) are shown for strange baryon to pion ratios in the top panels, and for anti-particles in the lower panels. 
The $p/\pi$ and $\bar{p}/\pi$ ratios have been included for comparison \cite{ref:Fuqiang}. 

It is seen that the strange baryon to pion ratio increases with centrality and the relative enhancement is higher than observed
in the $p/\pi$ ratio.
The multi-strange baryon $\Xi$ is the particle with the highest enhancement. 
In the $62.4$~GeV data, there are indications of a  difference in the enhancement rate between the strange baryons and  anti-baryons, with the latter showing a smaller increase with centrality.
This difference can be due to the non-zero net-baryon density which makes it easier to form strange baryons than strange anti-baryons. 
In addition, the lower net baryon density at $200$~GeV reduces the difference between strange baryons and anti-baryons at this energy. 
At $200$~GeV, there is no significant difference between baryon  and anti-baryon enhancement for all particle species, but the hierarchy of enhancement observed at 64.2 GeV is maintained at the top RHIC energy.
From the traditional strangeness enhancement data, where the yields are normalized by the $p+p$ data, the onset of the enhancement is pronounced and is already present in the most peripheral heavy-ion collisions, consistent with the effect that would be expected from the canonical suppression hypothesis.
In this context, the increase observed in Fig.~\ref{fig:BarionvsCentr}, which is normalized by the peripheral bin of the measured Au+Au collisions, does not include the effects of the canonical suppression and could be exclusively due to the enhancement of the $s\bar{s}$ pair production mechanism. 
However, the $p/\pi$ ratio also increases with the system size, reflecting the increase of the baryon stopping as the collisions becomes more central. 
This increase of the net-baryon density would also favor the production of $\Lambda$ and $\Xi$ with increasing centrality. 

Our results show that there is a net increase of the strange baryon production with respect to the pion yields as the system goes from peripheral to central collisions, suggesting a true increase of the $s\bar{s}$ pair production relative to the light quark production in central collisions. This conclusion is further strengthened by the observation of the enhancement in the $\phi / \pi$ ratio measured in central Au+Au collisions at $62.4$~GeV with respect to peripheral collisions as presented in Ref.~\cite{ref:STAR_PRC79_2009_064903,ref:phiPaper}. 

% Fig13
\begin{figure}[hbt]
 \includegraphics[width=8.5cm]{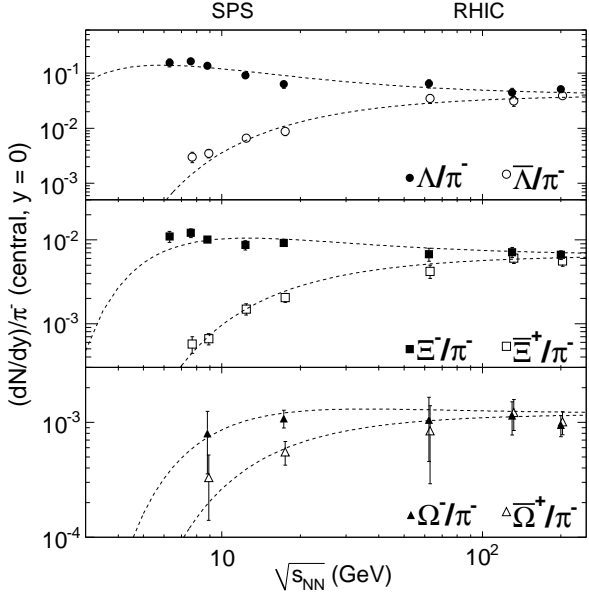} 
 \caption{Ratio of baryon (solid symbols) and anti-baryon (open symbols) to $\pi^{-}$ as a function of $\sqrt{s_{NN}}$. The lines are the results of the thermal model calculation (see text section \ref{sec:Chemical}). The SPS values are from NA49 \cite{ref:NA49_PRL93_022302_2004} (centrality $0-7$ \%) and the RHIC values are from STAR \cite{ref:STAR_PRL89_2002_092301, ref:STAR_PRL98_2007_062301} (centrality $0-5$ \%). For the multi-strange baryons $\Xi$ and $\Omega$ (bottom panel), the SPS results are from NA57 \cite{ref:NA57_PLB595_2004_68} (centrality $0-11$ \%) and the RHIC values are from STAR~\cite{ref:STAR_PRL92_2004_182301,ref:STAR_PRL98_2007_062301} (centrality $0-20$ \%).}
  \label{fig:BarionvsE}
\end{figure}

Figure~\ref{fig:BarionvsE} shows the ratios of strange baryons to $\pi$ meson yields as a function of collision energy. 
The anti-baryon to $\pi$ meson ratios increase as a function of energy while the $\Lambda/\pi$ and $\Xi/\pi$ ratios show a decrease from SPS to RHIC energies and then remain constant in the RHIC energy region.
The trend in the energy dependence of the $\Lambda/\pi$ ratio was already observed in the $K^{+}/\pi^{+}$ ratio where the peak structure at the SPS energies is more pronounced~\cite{ref:Fuqiang}. 
This shape is consistent with the net-baryon density dependence on the collision energy \cite{ref:Fuqiang}. 
The $\bar{\Lambda}/\pi$ and $\bar{\Xi}^{+}/\pi$ ratios show a strong monotonic increase with energy. 
The $\Omega^{-}/\pi$ and $\bar{\Omega}^{+}/\pi$ ratios show a smaller increase with energy, however the experimental uncertainties are large. 
The dashed lines in this figure correspond to the ratios of these particles using a thermal model calculation with parameters (temperature and chemical potentials) given by  the systematic study presented in Ref.~\cite{ref:Cleymans06}. 
The general trend is well reproduced by this calculation.

%============================================================
\subsection{Nuclear Modification Factors}
\label{subsec:Rcp}
%============================================================

One of the most important results from RHIC was the suppression of the particle spectra in the high-$p_{T}$ region observed via the nuclear modification factor known as $R_{AA}$. 
It is the ratio between the measured particle spectra from Au+Au collisions normalized by the corresponding number of binary collisions and the measured spectra in elementary $p+p$ collisions~\cite{ref:STAR_PRL_91_172302}. 
Together with the measurement of the disappearance of back-to-back high-$p_{T}$ hadron correlations in central Au+Au collisions attributed to interaction of the energetic jet particles with the formed medium~\cite{ref:STAR_PRL_90_082302}, it was concluded that matter created at RHIC is highly interacting.
The identified particle $R_{AA}$ from Au+Au collisions at $200$~GeV shows that even the strange particle spectra are also suppressed, and there is a grouping of baryons and mesons with respect to their suppression in the  intermediate-$p_{T}$ region.
This suppression seems to persist at lower energies and it was observed for pion, proton and charged kaons in Au+Au collisions at $62.4$~GeV~\cite{ref:STAR_PLB655_2007_104}.
It is important to study whether strange baryons follow the same trend as at $200$~GeV and whether the suppression also shows a baryon and meson grouping~\cite{ref:STAR_PRL_90_082302}.

% Fig14
\begin{figure}[h]
 \includegraphics[width=8.5cm]{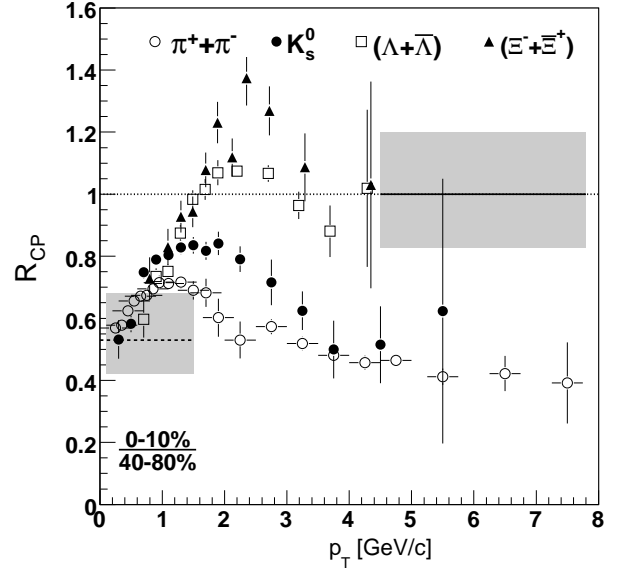}
 \caption{Nuclear modification factor $R_{CP}$, calculated as the ratio between 0-10\% central spectra and 40-80\% peripheral spectra, for $\pi$, $K^{0}_{S}$, $\Lambda$ and $\Xi$ particles in Au+Au collisions at $62.4$~GeV. The $\pi$ $R_{CP}$ values were extracted from Ref.~\cite{ref:STAR_PLB655_2007_104}. The gray band on the right side of the plot shows the uncertainties on the estimation of the number of binary collisions and the gray band on the lower left side indicates the uncertainties on the number of participants.}
 \label{fig:rcp62both}
\end{figure}

For the strange baryons measured at 62.4 GeV it is not possible to evaluate $R_{AA}$ due to the lack of $p+p$ data. 
However, a comparison between the spectra measured in central collisions and that from the peripheral collisions can also be used to analyse the suppression at high-$p_{T}$. 
This comparison is done with the differential nuclear modification factor $R_{CP}$ which is the ratio between the central and peripheral spectra scaled by the number of binary collisions ($\langle N_{\rm bin} \rangle $):

\begin{equation} R_{CP} = \left[ \frac{d^2 N^{\rm central} /dp_{T} dy}{d^2 N^{\rm peripheral}/dp_{T} dy} \right] \cdot\left[ \frac{N^{\rm peripheral}_{\rm bin}} {N^{\rm central}_{\rm bin}} \right] \end{equation}

The result of the $R_{CP}$ measurement comparing the spectra of central 0-10\% data with the peripheral 40-80\% data for the different particle species is shown in Fig.~\ref{fig:rcp62both}. 
For comparison, the $R_{CP}$ of pions from Au+Au collisions at $62.4$~GeV is also included~\cite{ref:STAR_PLB655_2007_104}.

The $K^{0}_{S}$ data, presented as solid circles show that the $R_{CP}$ reaches a maximum value of approximately 80\% at $p_{T} \approx 2$~GeV/c and decreases to a ratio of approximately 40\% for higher transverse momentum. 
The general shape of the $K^{0}_{S}$ $R_{CP}$ curve shows the same trend as the $\pi$ $R_{CP}$ (shown as open circles). 
In the intermediate-$p_{T}$ region, between 1.5 and 4~GeV/c, the $K^{0}_{S}$ ratio is higher than the $\pi$ ratio. 
The $\pi$ $R_{CP}$ reaches a maximum around 0.7 at a $p_{T}$ value of approximately 1.5~GeV/c. 
But, in the high-$p_{T}$ region, both $K^{0}_{S}$ and $\pi$ seem to be equally suppressed, saturating at around 40\%. 
The difference between the $K^{0}_{S}$ and $\pi$ curves at intermediate-$p_{T}$ is also observed in Au+Au collisions at $200$~GeV~\cite{ref:STAR_PRL_99_112301}.

% Fig15
\begin{figure}[h]
 \includegraphics[width=8.5cm]{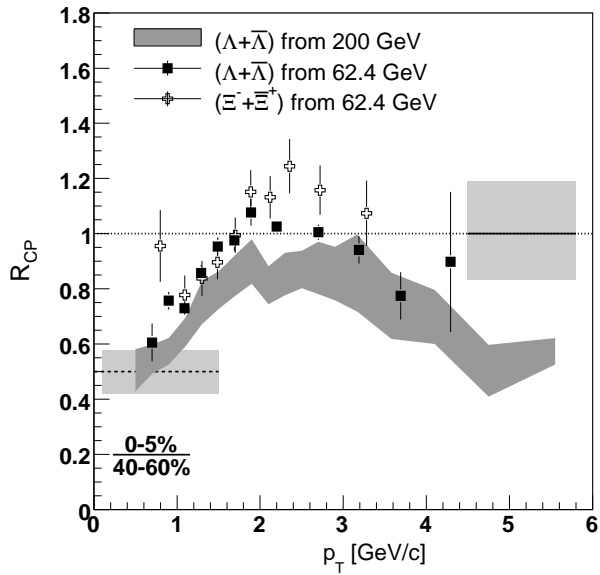}%
 \caption{Nuclear modification factor $R_{CP}$, calculated as the ratio between 0-5\% central spectra and 40-60\% peripheral spectra, for $\Lambda$ and $\Xi$ particles measured in Au+Au collisions at $62.4$~GeV. The gray band corresponds to the equivalent $R_{CP}$ curve for the $\Lambda$ particles measured in Au+Au collisions at $200$~GeV~\cite{ref:STAR_PRL98_2007_062301}.}
 \label{fig:rcp62and200both}
\end{figure}

The strange baryons, $\Lambda$ (open squares) and $\Xi$ (solid triangles), show a different behavior. The ratio increases to a maximum above unity at approximately $2.5$~GeV/c and then seems to decrease towards unity without showing significant suppression up to $4.5$~GeV/c. 
This behavior is different from that of the proton $R_{CP}$ which reaches unity at around $2$~GeV/c and starts to decrease for  higher values of $p_{T}$ \cite{ref:STAR_PLB655_2007_104}. 

In Fig.~\ref{fig:rcp62and200both}, the $R_{\rm CP}$ curves of $\Lambda$ and $\Xi$ measured in Au+Au collisions at $62.4$~GeV are compared to the $R_{\rm CP}$ of $\Lambda$ from Au+Au collisions at $200$~GeV presented in Ref.~\cite{ref:STAR_PRL98_2007_062301}.
To allow for a direct comparison to the $200$~GeV data, the $R_{CP}$ was calculated from the ratio between spectra of centrality bins 0-5\% and 40-60\%. 
The shape of the $R_{\rm CP}$ curve for $200$~GeV is similar to the results from 62.4 GeV, also showing a flat region after reaching the maximum, and the onset of suppression only occurs at higher values of $p_{T}$. 
The maximum in the intermediate-$p_{T}$ region seems to be higher at $62.4$~GeV than at $200$~GeV. 
However, considering the systematic uncertainty of the normalization factor, shown as
the gray band in the right side of Fig.~\ref{fig:rcp62and200both}, it is not possible to conclude that the ratio measured at $62.4$~GeV is higher than the ratio measured at $200$~GeV.

In summary, the $K^{0}_{S}$ spectra at $62.4$~GeV are consistent with a suppression scenario for $p_{T}>4.0$~GeV/c while the $\Lambda$ and $\Xi$ baryons seem to show no suppression for $p_{T} \approx 5$~GeV/c. 
In Fig.~\ref{fig:rcp62both} one observes that the difference between the $\Lambda$ and $\Xi$ curves is smaller than the gap to the $K^{0}_{S}$ data in the intermediate-$p_{T}$ region, suggesting a baryon-to-meson separation. 
However, at 62.4 GeV the baryon meson separation is not as clear as observed in the results from Au+Au collisions at $200$~GeV \cite{ref:STAR_PRL98_2007_062301}.

%============================================================
\subsection{Baryon to Meson Ratio}
\label{subsec:BaryonEnhancement}
%============================================================

Another interesting result related to particle spectra and yields obtained from RHIC data is the behavior of the proton to pion ratio with $p_{T}$, where an anomalous increase of baryon over meson production was observed in the intermediate-$p_{T}$ region and furthermore this increase was higher in more central collisions~\cite{ref:STAR_PRL_97_152301}. 
This result is referred to as the baryon anomaly, and it is the focus of theoretical discussions that include particle production and also particle interaction with the medium. 
A baryon-to-meson enhancement was also observed in the strangeness sector, where the ratio $\Lambda$ over $K^{0}_{S}$ also showed an anomalous increase in the intermediate-$p_{T}$ region \cite{ref:ProccedingMatt}.

\begin{figure}[h]
 \includegraphics[width=8.5cm]{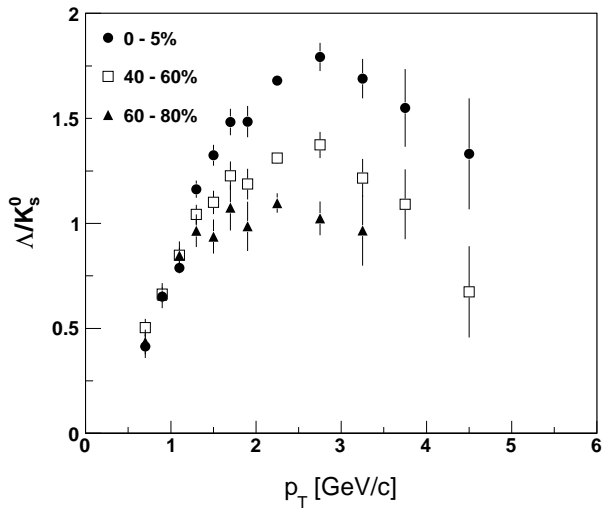}%
 \caption{$\Lambda$ / K$^0_S$ ratio as a function of transverse momentum for different centrality classes: $0-5$\% (solid circles), $40-60$\% (open squares) and $60-80$\% (solid triangles) in Au+Au collisions at $62.4$~GeV.}
\label{fig:LambdaK0s}
\end{figure}

Figure~\ref{fig:LambdaK0s} shows the centrality dependence of the $\Lambda$ to $K^{0}_{S}$ ratio as a function of $p_{T}$.
The ratio increases as a function of $p_{T}$ reaching a maximum around $2.5$~GeV/c. 
Above this $p_{T}$ there is a slope change in the curve, and the relative meson production starts to increase with respect to the baryons. 
In the intermediate-$p_{T}$ region, between $1.5$~GeV/c and $4$~GeV/c, there is a dominance of baryon over meson production.
Also, the maximum value of the ratios diminishes when going from central to more peripheral collisions.
One of the proposed theoretical explanations for this behavior involves the parton coalescence and recombination mechanisms in hadron production~\cite{ref:coalescence1,ref:coalescence2,ref:coalescence3}, where a baryon at a certain $p_{T}$ results from coalescing partons of a lower $p_{T}$. 
In this scenario, the baryon spectra would have a higher average $p_T$ than mesons, resulting in a ratio that increases with $p_{T}$.
In general, the coalescence models have been successful in describing the increase of the baryon-to-meson ratio, including the ratio with strange
particles.

% Fig16

An alternative proposal comes from the consideration of QCD higher-twist contributions in the initial state of hadron production~\cite{ref:Brodsky2008}, where baryons would be produced in a color singlet configuration, and thus would be color transparent and not suffer final state interactions, increasing the final yield of baryons over mesons.
Within this proposal, the increase of the baryon-to-meson ratio with the increase of centrality would be naturally explained by the higher degree of meson scattering as the volume of the medium increases.

% Fig17
\begin{figure}[h]
 \includegraphics[width=8.5cm]{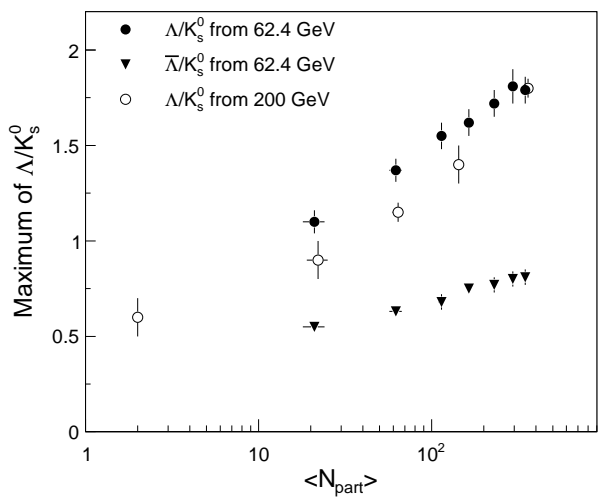}%
 \caption{Maximum value of the $\Lambda$ / K$^0_S$ ratio from Au+Au collisions at $62.4$~GeV (solid circles) and $200$~GeV (open circles) \cite{ref:ProccedingMatt} as a function of $\langle N_{\rm part} \rangle$ for different centrality classes. The lowest $\langle N_{part} \rangle$ point corresponds to $p+p$ collisions at $200$~GeV \cite{ref:STAR_pp}. The maximum of the $\overline{\Lambda}$ / K$^0_S$ from Au+Au collisions at $62.4$~GeV is shown as solid triangles.}
\label{fig:MaxLambdaK0s}
\end{figure}

In the high-$p_{T}$ region, the baryon-to-meson ratio starts to decrease with the onset of the contribution from jet fragmentation, in which meson dominates over baryon production.

In order to discriminate between different theoretical models it is necessary to 
study the systematic behavior for different energies and different system size.
Figure~\ref{fig:MaxLambdaK0s} shows the maximum value of the measured $\Lambda / K^{0}_{S}$ ratio for the different centrality bins (quantified here by the average number of participants) in Au+Au collisions at $62.4$~GeV and $200$~GeV \cite{ref:ProccedingMatt} as well as $p+p$ at $200$~GeV \cite{ref:STAR_pp}.
It is clear that the baryon-to-meson ratio exhibits the same trend at $62.4$~GeV as observed at $200$~GeV: an increase of the maximum with the increase of the centrality and the same ratio of approximately $1.7$ and $1.0$ for the most central and peripheral bins, respectively.
Figure~\ref{fig:MaxLambdaK0s} also shows the $\bar{\Lambda} / K^{0}_{S}$ ratio for the $62.4$~GeV data. The anti-baryon to meson ratio peaks at
much lower value, below unity, but shows the same trend of increase with centrality. The difference observed between the $\bar{\Lambda}$ and $\Lambda$
enhancement over the $K^{0}_{S}$ can be attributed to the non-zero net-baryon density.

%==================================================
\section{Conclusions}
\label{sec:conclusion}
%--------------------------------------------------
%==================================================

In this paper we present the strangeness production in Au+Au collisions at $\sqrt{s_{NN}}=62.4$~GeV as measured by the STAR detector at RHIC. 
For each particle and anti-particle which includes $K^{0}_{S}$, $\Lambda$, $\Xi$ and $\Omega$, the yield and $p_T$ spectra at mid-rapidity have been extracted for different collision centrality classes from peripheral to the most central events. 

Analyzing the anti-baryon to baryon ratio of the studied particles as a function of $\sqrt{s_{NN}}$ demonstrates that the mid-rapidity baryon-free regime is 
approached smoothly from the AGS energy to the top RHIC energy. 
The data at $\sqrt{s_{NN}}=62.4$~GeV presented here fit well into the systematics of the excitation function of these ratios. 
Also the production rates of all three anti-baryons studied show a smooth increase with energy whereas the baryon yield excitation functions are susceptible to the interplay between the increase of the pair production mechanism at mid-rapidity and the decrease of the net-baryon density with increasing rapidity range in higer energy collisions.

The chemical composition of the formed system was studied in the framework of a statistical thermal model, where the strangeness saturation factor 
parameterizes the degree of equilibration between the strange quarks and the lighter $u$ and $d$ quarks. 
Results using this model to fit our data show that even for the lower energy $62.4$~GeV data, the system created in central Au+Au collisions is still consistent with a saturation of strangeness production, similar to the result obtained at $200$~GeV.

Another similarity with the measurements at $200$~GeV is seen in the behavior in the intermediate-$p_{T}$ region. 
Results of the baryon-to-meson ratio and the nuclear modification factor point to a different behavior of mesons and baryons in this $p_{T}$ region. 
Various theoretical proposals were discussed which try to explain these experimental observations. 
To test any of these models, or even to place tighter constraints on the models, it is necessary to 
perform a careful and systematic comparison over a wide range of energies.
Within this context, the results presented here are a valuable complement to the existing systematics and fill the large gap between the lower
energy SPS data and the top RHIC data.

As an overall conclusion, it seems that strangeness production in Au+Au collision at $\sqrt{s_{NN}}=62.4$~GeV is qualitatively very similar in all aspects to the production at $\sqrt{s_{NN}}=200$~GeV. 
All excitation functions vary smoothly from AGS to top RHIC energies.

\vspace{-0.5cm}
%==================================================
\begin{acknowledgments}
We thank the RHIC Operations Group and RCF at BNL, the NERSC Center at LBNL and the Open Science Grid consortium for providing resources and support. This work was supported in part by the Offices of NP and HEP within the U.S. DOE Office of Science, the U.S. NSF, the Sloan Foundation, the DFG cluster of excellence `Origin and Structure of the Universe'of Germany, CNRS/IN2P3, FAPESP CNPq of Brazil, Ministry of Ed. and Sci. of the Russian Federation, NNSFC, CAS, MoST, and MoE of China, GA and MSMT of the Czech Republic, FOM and NWO of the Netherlands, DAE, DST, and CSIR of India, Polish Ministry of Sci. and Higher Ed., Korea Research Foundation, Ministry of Sci., Ed. and Sports of the Rep. Of Croatia, and RosAtom of Russia.

\end{acknowledgments}


\begin{thebibliography}{99}

\bibitem{ref:E896_Caines_JPG27_2001_311} H. Caines {\em et al.} (E896 Collaboration), J. Phys. G {\bf 27}, 311 (2001).
\bibitem{ref:NA57_PLB595_2004_68} F. Antinori {\em et al.} (NA57 Collaboration), Phys. Lett. B {\bf 595}, 68 (2004).
\bibitem{ref:NA49_PRL} C. Alt {\em et al.} (NA49 Collaboration), Phys. Rev. Lett. {\bf 94}, 192301 (2005).
\bibitem{ref:STAR_PRL89_2002_092301} C. Adler {\em et al.} (STAR Collaboration), Phys. Rev. Lett. {\bf 89}, 092301 (2002).
\bibitem{ref:Rafelski_2} J. Rafelski and B. Muller, Phys. Rev. Lett. {\bf 48}, 1066 (1982).
\bibitem{ref:NA35} J. Bartke {\em et al.} (NA35 Collaboration), Z. Phys. C {\bf 48}, 191 (1990).
\bibitem{ref:NA57} F. Antinori et al. (WA97/NA57 Collaboration), Nucl. Phys. A {\bf 698}, 118c (2002).
\bibitem{ref:WA85} Di Bari D. {\em et al.} (WA 85 Collaboration), N. Phys. A {\bf 590}, 307 (1995).
\bibitem{ref:HelenSTAR} B. I. Abelev {\em et al.} (STAR Collaboration), Phys. Rev. C {\bf 77}, 044908 (2008).
\bibitem{ref:STAR_PLB655_2007_104} B. I. Abelev {\em et al.} (STAR Collaboration), Phys. Lett. B {\bf 655}, 104 (2007).
\bibitem{ref:ProccedingMatt} M. A. C. Lamont, J. Phys. G {\bf 30}, S963-S968 (2004).
\bibitem{ref:CriticalPoint1} F. Karsch {\it et al.}, Nucl. Phys. Proc. Suppl. {\bf 129}, 614 (2004).
\bibitem{ref:CriticalPoint2} R. V. Gavai and S. Gupta, Phys. Rev. D {\bf 71}, 114014 (2005).
\bibitem{ref:Fuqiang}B. I. Abelev {\em et al.} (STAR Collaboration), Phys. Rev. C {\bf 79}, 034909 (2009).
\bibitem{ref:STAR_PRL98_2007_062301} J. Adams {\em et al.} (STAR Collaboration), Phys. Rev. Lett. {\bf 98}, 062301 (2007).
\bibitem{ref:STARNIM} K.H. Ackermann {\em et al.} (STAR Collaboration), Nucl. Instrum. and Meth. A {\bf 499}, 624 (2003).
\bibitem{ref:NIMPID} M. Shao, O. Barannikova, X. Dong, Y. Fisyak, L. Ruan, P. Sorensen, Z. Xu, Nucl. Instrum. and Meth. A {\bf 558}, 419 (2006).
\bibitem{ref:STARTrigger} F.S. Bieser {\em et al.}, Nucl. Instrum. and Meth. A {\bf 499}, 766 (2003).
\bibitem{ref:Glauber} M. L. Miller, K. Reygers, S. J. Sanders and P. Steinberg, Annu. Rev. Nucl. Part. Sci. {\bf 57}, 205 (2007). 
\bibitem{ref:NA49_PRL93_022302_2004} T. Anticic {\em et al.} (NA49 Collaboration), Phys. Rev. Lett. {\bf 93}, 022302 (2004).
\bibitem{ref:STAR_PRL92_2004_182301} J. Adams {\em et al.} (STAR Collaboration), Phys. Rev. Lett. {\bf 92}, 182301 (2004).
\bibitem{ref:Cleymans_all}J. Cleymans, H. Oeschler, K. Redlich and S. Wheaton, Phys. Rev. C {\bf 73}, 034905 (2006).
\bibitem{ref:Cleymans} J. Cleymans, B. Kaempfer, S. Wheaton, Phys. Rev. C {\bf 65}, 027901 (2002).
\bibitem{ref:Becattini} F. Becattini, M. Gazdzicki, A. Keranen, J. Manninen, R. Stock, Phys. Rev. C {\bf 69}, 024905 (2004).
\bibitem{ref:PBM} P. Braun-Munzinger, D. Magestro, K. Redlich, J. Stachel, Phys. Lett. B {\bf 518}, 41 (2001).
\bibitem{ref:Rafelski} J. Rafelski, J. Letessier, G. Torrieri, Phys. Rev. C {\bf 72}, 024905 (2005).
\bibitem{ref:Kaneta_Xu} N. Xu and M. Kaneta, Nucl. Phys. A {\bf 698}, 306 (2002).
\bibitem{ref:Jun_1} J. Takahashi (for the STAR Collaboration), J. Phys. G {\bf 36}, 064074 (2009).
\bibitem{ref:PBM_1999} P.~Braun-Munzinger, I. Heppe and J. Stachel, Phys. Lett. B {\bf 465}, 15 (1999).
\bibitem{ref:Cleymans06} J. Cleymans, H. Oeschler, K. Redlich, S. Wheaton, J. Phys. G {\bf 32}, S165 (2006). 
\bibitem{ref:THERMUS} S. Wheaton, J. Cleymans, M. Hauer, Computer Physics Communications {\bf 180}, 84 (2009).
\bibitem{ref:Jun02} J. Takahashi (for the STAR Collaboration), J. Phys. G {\bf 35}, 044007 (2008).
\bibitem{ref:Redlich} A. Tounsi, A. Mischke and K. Redlich, Nucl. Phys. A {\bf 715}, 565 (2003).
\bibitem{ref:STAR_PRC79_2009_064903} B. I. Abelev {\em et al.} (STAR Collaboration), Phys. Rev. C {\bf 79}, 064903 (2009).
\bibitem{ref:phiPaper} B. I. Abelev {\em et al.} (STAR Collaboration), Phys. Lett. B {\bf 673},  183 (2009).
\bibitem{ref:STAR_PRL_91_172302} J. Adams {\em et al.} (STAR Collaboration), Phys. Rev. Lett. {\bf 91}, 172302 (2003).
\bibitem{ref:STAR_PRL_90_082302}  C. Adler {\em et al.} (STAR Collaboration), Phys. Rev. Lett. {\bf 90}, 082302 (2003).
\bibitem{ref:STAR_PRL_99_112301} B. I. Abelev {\em et. al.} (STAR Collaboration), Phys. Rev. Lett.  {\bf 99}, 112301 (2007).
\bibitem{ref:STAR_PRL_97_152301}  B. I. Abelev {\em et al.} (STAR Collaboration),  Phys. Rev. Lett. {\bf 97}, 152301 (2006).
\bibitem{ref:coalescence1} R. C. Hwa, C. B. Yang, Phys. Rev. C {\bf 67}, 034902 (2003).
\bibitem{ref:coalescence2} V. Greco, C. M. Ko, P. Levai, Phys. Rev. Lett. {\bf 90}, 202302 (2003).
\bibitem{ref:coalescence3} R. J. Fries, B. Muller, C. Nonaka, S. A. Bass, Phys. Rev. C {\bf 68}, 044902 (2003).
\bibitem{ref:Brodsky2008} S.J. Brodsky, A. Sickles, Phys. Lett. B {\bf 668}, 111 (2008).
\bibitem{ref:STAR_pp}  B. I. Abelev {\em et al.} (STAR Collaboration), Phys. Rev. C {\bf 75}, 064901 (2007).
\end{thebibliography}
\end{document}